\documentclass{article}

\usepackage{arxiv}

\usepackage[linesnumbered,ruled,vlined]{algorithm2e}
\SetKwInput{KwInput}{Input}                
\SetKwInput{KwOutput}{Output}              
\SetCommentSty{mycommfont}
\usepackage{glossaries}
\newacronym{AD}{AD}{angle-delay}
\newacronym{AF}{AF}{angle-frequency}
\newacronym{AWGN}{AWGN}{additive white Gaussian
noise}
\newacronym{BICM-OFDM}{BICM-OFDM}{bit-interleaved coded \gls{OFDM}}
\newacronym{BICM}{BICM}{bit-interleaved coded}
\newacronym{BG}{BG}{Bernoulli Gaussian}
\newacronym{BP}{BP}{belief propagation}
\newacronym{BS}{BS}{base station}
\newacronym{B-TSGM}{B-TSGM}{Bernoulli two-state Gaussian mixture}
\newacronym{CE}{CE}{channel estimation}
\newacronym{CS}{CS}{compressed sensing}
\newacronym{CSI}{CSI}{channel state information}
\newacronym{DFT}{DFT}{discrete Fourier transform}
\newacronym{EP}{EP}{expectation propagation}
\newacronym{EPA}{EPA}{expectation propagation approximation}
\newacronym{EM}{EM}{expectation maximization}
\newacronym{FDD}{FDD}{frequency-division duplexing}
\newacronym{GAMP}{GAMP}{generalized approximate message passing}
\newacronym{HMP}{HMP}{hybrid message passing}
\newacronym{LMMSE}{LMMSE}{linear minimum mean square error}
\newacronym{MC}{MC}{Markov chain}
\newacronym{MF}{MF}{mean field}
\newacronym{MIMO}{MIMO}{multiple-input multiple-output}
\newacronym{MIMO-OFDM}{MIMO-OFDM}{multiple-input multiple-output orthogonal frequency-division multiplexing}
\newacronym{MP}{MP}{message passing}
\newacronym{MSE}{MSE}{mean square error}
\newacronym{NMSE}{NMSE}{normalized mean square error}
\newacronym{NNSPL}{NNSPL}{nearest neighbor sparsity pattern learning}
\newacronym{OFDM}{OFDM}{orthogonal frequency-division multiplexing}
\newacronym{PDF}{PDF}{probability density function}
\newacronym{PDFT-RP}{PDFT-RP}{partial DFT random permutation}
\newacronym{SCM}{SCM}{spatial channel model}
\newacronym{SE}{SE}{state evolution}
\newacronym{STF}{STF}{structured turbo framework}
\newacronym{SNR}{SNR}{signal-to-noise ratio}
\newacronym{STCS}{STCS}{structured turbo-compressed sensing}
\newacronym{STCS-DS}{STCS-DS}{STCS-delay support}
\newacronym{STCS-FS}{STCS-FS}{structured turbo-compressed sensing frequency support}
\newacronym{TSGM}{TSGM}{two-state Gaussian mixture}
\newacronym{TSGM-LVD}{TSGM-LVD}{two-state Gaussian mixture with large variance difference}
\newacronym{ULA}{ULA}{uniform linear array}
\newacronym{5G}{5G}{5th-generation}
\newacronym{6G}{6G}{6th-generation}

\usepackage{amsmath}
\usepackage{makecell}
\usepackage{diagbox}
\usepackage{amsmath,amssymb,amsfonts}
\usepackage[normalem]{ulem}
\usepackage{cite}
\usepackage{color}
\usepackage{multirow}
\usepackage{subcaption}
\usepackage{graphicx}
\usepackage{textcomp}
\usepackage{xcolor}
\usepackage{epstopdf}
\usepackage{stfloats}
\usepackage{float}
\usepackage{hyperref}
\hypersetup{pdfborder = {0 0 0}}
\usepackage{appendix}
\usepackage{tikz}
\newcommand*{\circled}[1]{\lower.7ex\hbox{\tikz\draw (0pt, 0pt)%
    circle (.5em) node {\makebox[1em][c]{\small #1}};}}
\definecolor{amaranth}{rgb}{0.9, 0.17, 0.31}
\newcommand{\revise}[1]{\textcolor{black}{#1}}

\begin{document}
\graphicspath{{Figure/}}
\title{Hybrid Message Passing Algorithm for Downlink FDD Massive MIMO-OFDM Channel Estimation}
\author{Yi Song, Chuanzong Zhang, Xinhua Lu, Fabio Saggese, Zhongyong Wang*}
\vspace{-1cm}

\thanks{Y. Song and Z. Wang are with the Department of Information Engineering, Zhengzhou University, 450001 Zhengzhou, China. Emails: songyizzu@gs.zzu.edu.cn, iezywang@zzu.edu.cn.}
\thanks{C. Zhang and X. Lu are with the Research Center for Communications and Signal Processing, Nanyang Institute of Technology, 473000  Nanyang, China. Emails: ieczzhang@gmail.com, ieluxinhua@sina.com.}
\thanks{F. Saggese is with the Department of Electronic Systems, Aalborg University, 9220 Aalborg, Denmark. Email: fasa@es.aau.dk.}

\maketitle
\begin{abstract}
The design of message passing (MP) algorithms on factor graphs is an effective manner to implement \revise{channel estimation (CE) in wireless communication systems, which performance can be further improved by exploiting prior probability models that accurately match the channel characteristics.
In this work, we study the CE problem in a downlink massive multiple-input multiple-output (MIMO) orthogonal frequency division multiplexing (OFDM) system. As the prior probability, we propose the Markov chain two-state Gaussian mixture with large variance differences (TSGM-LVD) model to exploit the structured sparsity in the angle-frequency domain of the channel. Existing single and combined MP rules cannot deal with the message computation of the proposed probability model. To overcome this issue, we present a general method to derive the hybrid message passing (HMP) rule, which allows the calculation of messages described by mixed linear and non-linear functions.
Accordingly, we design the HMP-TSGM-LVD algorithm under the structured turbo framework (STF).}
Simulation results demonstrate that the proposed algorithm \revise{converges faster and obtains better and more stable performance than its counterparts. In particular, the gain of the proposed approach is maximum (3 dB) in the high signal-to-noise ratio regime, while benchmark approaches experience oscillating behavior due to the improper prior model characterization.}
\end{abstract}

\keywords{Massive MIMO, OFDM, FDD channel estimation, TSGM-LVD channel model, message passing.}

\section{Introduction}
The combination of massive \gls{MIMO} and \gls{OFDM} techniques can achieve huge performance gains in both spectrum and energy efficiency, playing an important role in 5th-generation (5G) applications~\cite{OverviewMIMO,cellular5G} and being a focused research area for 6th-generation (6G) technologies~\cite{MIMO6G}.

Massive MIMO-OFDM \gls{CE} has been one of the main focuses of academia and industry in recent years. Accurate channel state information (CSI) is essential for signal detection, resource allocation and beamforming. The \gls{CE} problems in time division duplex (TDD) and frequency division duplex (FDD) are different. In TDD system, the CE problem can be solved by channel reciprocity property~\cite{TDDFDD}.
For now, most of the contemporary cellular networks have adopted the FDD protocol, which is more efficient for delay-sensitive and symmetric traffic applications~\cite{FDDdomainted}. However, in FDD massive MIMO-OFDM systems, it is challenging to improve the downlink CSI estimation accuracy without increasing the complexity of the algorithm and the pilot overhead.

Fortunately, massive MIMO-OFDM channels usually exhibit sparsity in the transform domain~\cite{tse2005s} due to the limited local scatterers in physical environments. Therefore, many compressed sensing (CS) based approaches have been proposed by exploiting sparsity information about the transform domain channel \cite{EBSLiuAn,rao2015compressive,han2017compressed, Gao2015FDD, Gao2016FDD}. However, applying the above CS-based algorithms to the FDD downlink channel estimation problem straightforwardly is limited in massive MIMO-OFDM systems.
Specifically, CS-based methods are susceptible to noise interference, while the observation matrix must satisfy the restricted isometry property (RIP).

Recently, \gls{MP} is receiving increasing attention from researchers due to its low computational complexity and high performance in terms of approximation~\cite{SPA,Donoho,Merging}. A number of prior works investigated CSI estimation problem with MP~\cite{BICMOFDM, BPMFstretched, yuanGAMPMF, TurboDFT, STCSMP}. 
In those works, MP tends to obtain better performance compared with traditional CS-based algorithms. More specifically, both the channel prior model which exploits the channel sparsity and message computation rules are critical aspects affecting the performance of MP algorithms employed for CE~\cite{STCSMP,BPMFstretched}. Thus, in this work we focus on channel prior model construction and message computation rules selection.

\subsection{Related Work}
Researchers have developed many solutions focused on channel prior models~\cite{BICMOFDM,STCSMP,ChenSTCS,STCSFS}. The authors in \cite{BICMOFDM} proposed a Markov chain two-state Gaussian mixture (TSGM) prior model to exploit channel-tap sparsity and cluster structure for bit-interleaved coded orthogonal frequency division multiplexing (BICM-OFDM) system. \cite{STCSMP} presented a Markov chain Bernoulli Gaussian (BG) prior model to design the structured turbo compressed sensing (STCS) algorithm by exploiting the clustered sparsity of the massive MIMO channel in the \gls{AF} domain. \cite{ChenSTCS} further extended the STCS algorithm to the massive MIMO-OFDM system. Employing the Markov chain BG prior model while proposing a structured turbo compressed sensing with frequency support (STCS-FS) algorithm, \cite{ChenSTCS}  achieved a considerably lower mean square error (MSE) performance compared with the CS-based algorithms with frequency support. Furthermore,~\cite{STCSFS} explored the massive MIMO-OFDM channel structured sparsity in the angle-delay domain, combined with Markov chain BG prior model to design structured turbo compressed sensing with delay support (STCS-DS) algorithm, which has the state-of-the-art performance in both complexity and convergence speed.

All of the above works~\cite{STCSMP, ChenSTCS, STCSFS} utilized the Markov chain BG model to characterize the massive MIMO-OFDM channels. The BG model assumes that the non-zero elements of the channel follow a Gaussian distribution while its small value elements are zero. On the contrary, in this work, we show that the elements with relatively small values of the channel in the \gls{AF} domain are not zero but close to zero, coherently labelled as ``\emph{near-zero}'' elements. 
Accordingly, the TSGM model~\cite{BICMOFDM} is more appropriate, making use of a Gaussian distribution for near-zero elements. 
However, TSGM assumes that the variances of the Gaussian distributions are the same. We find that the values of the non-zero elements within the \gls{AF} domain channel vary significantly. According to the physical channel characteristics, it is more accurate to set different variances for the non-zero elements, while the near-zero elements' Gaussian distributions have the same variance value. 
Based on the above considerations, we propose the two-state Gaussian mixture with large variance differences (TSGM-LVD) prior model to further characterize the near-zero elements and the large variation of the non-zero elements of the massive MIMO-OFDM channel. 
Furthermore, \cite{BICMOFDM, STCSMP, ChenSTCS, STCSFS} utilized the expectation maximization (EM) algorithm to update the hyperparameters of BG, TSGM, and the Markov chain. Differently from them, we treat TSGM-LVD hyperparameters as variables, updating them using MP algorithms. Indeed, updating the hyperparameters on the factor graph can simplify the derivation of the EM. Moreover, designing a suitable prior distribution for the hyperparameters can improve the algorithm convergence speed without increasing the complexity. 

Constructing a prior model which accurately characterizes the physical channel is a prerequisite for achieving high accuracy CSI. Nevertheless, the challenge of designing a high-performance MP algorithm lies also in selecting the appropriate message calculation rules.
In literature, three \emph{single} calculation rules are commonly used to implement MP algorithms: \gls{BP}, also known as sum-product algorithm~\cite{SPA}, \gls{MF}, also called variational message passing (VMP)~\cite{VMP}, and \gls{EP}~\cite{EP}. These three rules have their specific application scenarios: \gls{BP} rule is well-suited for discrete models with hard constraints and linear Gaussian models, while the performance deteriorates when applied to multi-variable product calculations or non-linear models~\cite{BPMFstretched}; \gls{MF} rule is especially useful for the estimation of continuous parameters (e.g., noise variance), but it may exhibit poor performance when applied to multi-variable summation calculations or discrete variables with hard constraints~\cite{YuanBPMF}; \gls{EP} rule can be regarded as an approximation of the \gls{BP} rule, where beliefs are approximated by distributions in a specific exponential family~\cite{Mihai}. In general, it is difficult to use a single rule to compute all the message calculations due to multiple types of variables and the complicated relationship among them. 
To address this issue,~\cite{Merging} proposed the \emph{combined} BP-MF rule, used as a base for a considerable number of works~\cite{Mihai, BPMFstretched, yuanGAMPMF, YuanBPMF}.
However, faced with \emph{mixed linear and non-linear models}, i.e., in the presence of complicated factor nodes expressing joint product summation operations during message computation, the combined rule can not be applied directly; indeed, the aforementioned combined rule is constrained to use \gls{BP} \emph{or} MF at each node, not being able to deal with mixed models. Unfortunately, these mixed linear and non-linear models are common in practical communication systems.

\revise{Recently,~\cite{HMPrule} proposed a framework based on \emph{constrained Bethe free energy} (BFE) minimization approach; BFE is used to derive, among others, a \gls{HMP} rule through the formulation of different constraints on system variables. The main advantage of the \gls{HMP} rule is that enables the simultaneous use of different single message calculation rules at the same factor node, allowing the evaluation of messages in presence of mixed linear and non-linear probability models, overcoming the main limitation of single and combined rules. Based on~\cite{HMPrule},~\cite{HMPapplication} applied the BFE minimization framework to solve the massive MIMO \gls{CE} problem, implicitly applying \gls{HMP} rule in the derivation of the algorithm. 
However,~\cite{HMPrule,HMPapplication} focus on the BFE framework and do not directly utilize an \gls{HMP} approach; the main disadvantage is the need of reformulating and solving the constrained BFE minimization problem any time we deal with a new problem. This prevents the direct use of the \gls{HMP} rule on the factor graph which is a more simpler and straightforward approach.}

\revise{Differently from other works in literature, in this article, we propose a generalized formulation and derivation of the \gls{HMP} rule that allows us to design \gls{MP} algorithms omitting the BFE derivation process. Furthermore, we apply the proposed \gls{HMP} rule derivation on the proposed \gls{TSGM-LVD} prior model, developing a \gls{STF} \gls{CE} algorithm for downlink massive \gls{MIMO}-\gls{OFDM} scenario. A detailed list of contribution is given in the next subsection.}
We summarize the characteristics of various models and message calculation rules in Table~\ref{tab:evolution}; the proposed TSGM-LVD model using the HMP rule has clear advantages over the others, as we will demonstrate throughout the article.

\revise{Finally, we remark that the \gls{HMP} rule studied in this article is different from most of the existing works related to \gls{HMP} approaches~\cite{Cuilocalization, HMPidea, HMPapproachDai}. The \gls{HMP} mentioned in~\cite{Cuilocalization, HMPidea} refers to approximation methods aiming to reduce the complexity of message computations, while~\cite{HMPapproachDai} refers to the iterative exchange of messages between two modules performing different estimation processes. Instead, the \gls{HMP} rule proposed in this article is a new method of message computation for factor graphs, being an evolved version of single and combined computation rules.} 

\begin{table}[h]
	\centering
	\caption{Summary of related channel prior models and MP rules.}
	\label{tab:evolution}  
	{\scriptsize
	\begin{tabular}{c|c|c|c} 
		\hline 
		\diagbox[width=2.2cm,height=1.4cm]{Rule}{Prior model}
		& BG\cite{STCSFS}
		& TSGM\cite{BICMOFDM}
		& TSGM-LVD (proposed)
		\\
		\hline
		BP-EM\cite{STCSFS}
		& \makecell[l]{Neglects the near-zero elements\\ Parameters updated by EM}
		& \makecell[l]{Neglects the variation of non-zero elements\\  Parameters updated by EM} 
		& $-$
		\\
		\hline
	    HMP\cite{HMPrule}
	    & \makecell[l]{Neglects the near-zero elements\\ Parameters treated as variables}
	    & \makecell[l]{ Neglects the variation of non-zero elements \\ Parameters treated as variables}
	    &\makecell[l]{ Consider channel full characteristics 
	    \\Parameters treated as variables}
        \\
		\noalign{\smallskip}\hline
	\end{tabular}}
\end{table}

\subsection{Main Contributions}
\revise{The main contributions of this article can be summarized as follows:}
OFDM FDD downlink CE problem modeled as a TSGM-LVD. 
\begin{itemize}
  \item \textbf{Generalized hybrid message passing rule}: \revise{First, we propose a general formulation of the \gls{HMP} rule and its derivation. In this way, we can apply the \gls{HMP} rule to design MP algorithms directly on the factor graph and omit the BFE derivation process. Second, we innovatively divide the factor graph from the edge perspective. The above classification method enables the application of different message updating rules to different messages departing, effectively handling the problem of message computation for mixed linear and non-linear models. Third, we give new insight into the connection among the \gls{HMP} rule, the single \gls{MP} rules and the combined \gls{MP} rule from the edge perspective.}
  \item \textbf{TSGM-LVD probability model}: \revise{We study the \gls{AF} domain channel characteristics of the massive MIMO-OFDM system and propose the \gls{TSGM-LVD} probability model, which can fully exploit the clustered sparsity of \gls{AF} domain channel. As a baseline, we utilize the \gls{TSGM} probability model to indicate the numerical characteristics of the non-zero and near-zero elements of the \gls{AF} domain channel. On top of that, we propose the \gls{TSGM-LVD} probability model to account for the large variation among non-zero elements' channel. Finally, we use the Markov chain formulation to represent the cluster sparsity of non-zero and near-zero elements within the \gls{AF} domain channel.}
  \item \textbf{Hybrid message passing algorithm for massive MIMO-OFDM CE}: \revise{Based on the TSGM-LVD probability model, we utilize the proposed \gls{HMP} rule to design the \gls{AF} domain \gls{CE} algorithm for massive MIMO-OFDM system, yielding the HMP-TSGM-LVD algorithm, whose convergence can be well predicted by the \gls{SE}. Numerical examples show that the proposed HMP-TSGM-LVD algorithm exhibits the fastest convergence, and provides better \gls{NMSE} performance than the state-of-the-art STCS-FS algorithm~\cite{STCSMP,ChenSTCS,STCSFS} while maintaining the same complexity.}
\end{itemize}

The rest of the article is organized as follows. In Section~\ref{sec:HMP}, we present the HMP rule, whose detailed derivation is given in the Appendix A. Then, we present the system model and the system factor graph in Section~\ref{sec:model}. In Section~\ref{sec:BPHF}, we propose the HMP-TSGM-LVD channel estimation algorithm for massive MIMO-OFDM system. Simulation results are given in Section~\ref{sec:Sim}. Our conclusions are finally drawn in Section~\ref{sec:Conclusion}.

\emph{Notation}: Boldface lowercase and uppercase letters denote vectors and matrices, respectively; superscripts $(\cdot )^{\mathrm{T}}$ and $(\cdot )^{\mathrm{H}}$ denote transposition and Hermitian transposition, respectively. $|\mathcal{I}|$ denotes the cardinality of a finite set $\mathcal{I}$, the relative complement of $\left\{i\right\}$ in $\mathcal{I}$ is written as $\mathcal{I}\setminus i$. The expectation operator of a function $f(x)$ with respect to a probability density function (PDF) $g(x)$ is expressed as $\mathbb{E}[f(x)]_{g(x)}=\left \langle f(x)\right \rangle _{g(x)}=\int f(x)g(x)\mathrm{d}x/\int g(x)\mathrm{d}x$; $\mathbb{V}\mathrm{ar} (x)_{g(x)}=\left \langle \left |x\right|^2\right\rangle _{g(x)}- |\left \langle x\right \rangle_{g(x)}|^2 $ stands for the variance. We denote that a variable $x$ follows a complex Gaussian distribution with mean $\mu$ and variance $v$ using $\mathcal{CN} (x;\mu ,v)$. $\mathrm{Ga}(\cdot;a,b)$ denotes a Gamma PDF with shape parameter $a$ and rate parameter $b$. $\mathrm{Beta}(\cdot;e,f)$ denotes a Beta PDF with two shape parameters $e$ and  $f$. The relation $f(x) = cg(x)$ for some positive constant $c$ is written as $f(x) \propto g(x)$.

\section{Hybrid Message Passing Rule}
\label{sec:HMP}
In this section, we introduce the combined BP-MF rule; based on this rule, we present a new approach to derive HMP rule~\cite{HMPrule} in a more general and intuitive way. Then, we propose a new approach to analyze the relationships between the HMP rule and single, combined BP-MF rules.

\subsection{Combined BP-MF rule}
Let $p(\boldsymbol{x})$ be an arbitrary PDF of a random vector $\boldsymbol{x}\triangleq (x_i|i\in \mathcal{I} )^{\mathrm{T}}$. We group all the factors represented by the set $\mathcal{A}$ into two disjunctive sets: $\mathcal{A}_{\mathrm{BP}}\cap \mathcal{A}_{\mathrm{MF}}=\oslash$ and $\mathcal{A}_{\mathrm{BP}}\cup  \mathcal{A}_{\mathrm{MF}}=\mathcal{A}$. After the factorization, the PDF can be written as
\begin{equation}
    \label{eq:BPMFFactorize}
    p(\boldsymbol{x})=\prod_{a\in \mathcal{A}}f_a(\boldsymbol{x}_a)=\prod_{b\in \mathcal{A}_{\mathrm{BP}}}f_b(\boldsymbol{x}_b)\prod_{c\in \mathcal{A}_{\mathrm{MF}}}f_c(\boldsymbol{x}_c),
\end{equation}
where $\boldsymbol{x}_a, \boldsymbol{x}_b$ and $\boldsymbol{x}_c$ denote the vector of the variables $x_i$ that are arguments of the factor nodes $f_a, f_b$ and $f_c$, respectively. Moreover, we define $\mathcal{N}(a)\subseteq \mathcal{I}$ to be the set of indices of all variables $x_i$ that are arguments of factor node $f_a$. Correspondingly, $\mathcal{N}(i) \subseteq \mathcal{A}$ denotes the set of indices of all factor nodes $f_a$ that depend on $x_i$. The parts of the factorization that correspond to ${\textstyle \prod_{b\in\mathcal{A}_{\mathrm{BP}}}} f_b(\boldsymbol{x}_b)$ and $\mathcal{I}_{\mathrm{BP}}\triangleq {\textstyle \bigcup_{b\in \mathcal{A}_{\mathrm{BP} }}}\mathcal{N}(b)$ are referred to as ``BP part''; $ {\textstyle \prod_{c\in\mathcal{A}_{\mathrm{MF}}}} f_c(\boldsymbol{x}_c)$ and $\mathcal{I}_{\mathrm{MF}}\triangleq {\textstyle \bigcup_{c\in \mathcal{A}_{\mathrm{MF} }}}\mathcal{N}(c)$ are referred to as ``MF part''. The combined BP-MF rule~\cite{Merging,Mihai} is
\begin{eqnarray}
    \label{eq:comBP}
    m_{f_a\to x_{i}}^{\mathrm{BP} }(x_{i}) &=& \int f_a\left(\boldsymbol{x}_a\right) \prod_{j\in \mathcal{N}{\left(a\right)}\setminus i}
    n_{x_{j}\to f_a}(x_{j})\mathrm{d}x_{j} ,\forall a\in \mathcal{A}_{\mathrm{BP}}, i\in \mathcal{N}{\left(a\right)},
    \\
    \label{eq:comMF}
     m_{f_a\to x_{i}}^{\mathrm{MF} }(x_{i}) &=& \mathrm{exp}\left \{\left\langle\mathrm{ln}f_a\left(\boldsymbol{x}_a\right )  \right\rangle_{\prod_{j\in \mathcal{N}\left(a\right)\setminus i} n_{x_j \to f_a}(x_{j})} \right \} ,\forall a\in \mathcal{A}_{\mathrm{MF}}, i\in \mathcal{N}\left(a\right),
    \\
    \label{eq:nmessage}
    n_{x_i\to f_a}(x_{i}) &\propto& \prod_{b \in \left(\mathcal{A}_{\mathrm{BP}}\cap \mathcal{N}\left(i\right)\right)\setminus a}m^{\mathrm{BP}}_{f_b\to x_i}\left (x_i\right ) \prod_{c \in \left(\mathcal{A}_{\mathrm{MF}}\cap \mathcal{N}\left(i\right)\right)}m^{\mathrm{MF}}_{f_c\to x_i}\left (x_i\right ), \forall i \in \mathcal{I}.
\end{eqnarray} 
The sets of $\mathcal{A}_{\mathrm{BP}}$ and $\mathcal{A}_{\mathrm{MF}}$ with the combined BP-MF rule can only connect variables of the corresponding type $\mathcal{I}_{\mathrm{BP}}$ and $\mathcal{I}_{\mathrm{MF}}$, respectively. Therefore, only one rule can be used at one specific factor node. However, there are some factor nodes expressing joint product and summation operations in the practical system,  on which the calculation cannot be completed using the combined BP-MF rule directly.

\subsection{Hybrid message passing rule}
\begin{figure}
    \begin{centering}
        \includegraphics[scale=0.8]{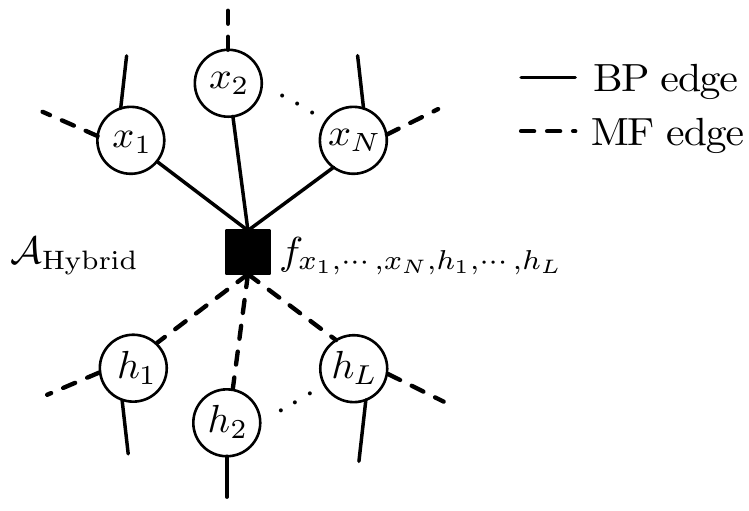}
        \caption{A partial factor graph representing the HMP rule.}
        \label{fig:HMP}
    \end{centering}
\end{figure}

In this subsection we present the HMP rule from the edge perspective. The factor graphs consist of variable nodes $x_i,i\in \mathcal{I}$, factor nodes $f_a,a\in \mathcal{A}$, and edges $ai\in \mathcal{E}$ connected with $f_a$ and $x_i$~\cite{SPA}. We group all edges $\mathcal{E}$ into two sets as $\mathcal{E}_{\mathrm{BP}}$ and $\mathcal{E}_{\mathrm{MF}}$, which satisfy $\mathcal{E}_{\mathrm{BP}}\cap \mathcal{E}_{\mathrm{MF}}=\oslash$ and $\mathcal{E}_{\mathrm{BP}}\cup  \mathcal{E}_{\mathrm{MF}}=\mathcal{E}$. All factor nodes are grouped into the set $\mathcal{A}_{\mathrm{Hybrid}}$. We define $\mathcal{N}_{\mathrm{BP}}(a)$ and $\mathcal{N}_{\mathrm{MF}}(a)$ as the sets of variable indices connected to factor node $f_a$ by BP edges and MF edges, respectively. Correspondingly, the sets of factor indices connected to $x_i$ by BP edges and MF edges can be expressed as $\mathcal{N}_{\mathrm{BP}}(i)$ and $\mathcal{N}_{\mathrm{MF}}(i)$, respectively. Unlike the derivation given in~\cite{HMPrule}, we have reformulated the HMP rule to make it more intuitive and efficient. The HMP rule is
\begin{eqnarray}
    \label{eq:HtoBP}
    m_{f_a\to x_{i}}^{\mathrm{BP}}(x_{i}) &=& \int \mathrm{exp}\Big \{\left\langle\mathrm{ln}f_a \right\rangle_{\prod_{j\in \mathcal{N}_{\mathrm{MF}}(a)} b(h_{j})} \Big \} \prod_{k\in \mathcal{N}_{\mathrm{BP}}(a)\setminus i}
    n_{x_{k}\to f_a}(x_{k})\mathrm{d}x_{k} ,\forall ai\in \mathcal{E}_{\mathrm{BP}}, 
     \\
    \label{eq:nBPHMP}
    n^{\mathrm{BP}}_{x_i\to f_a}(x_{i}) &\propto& \prod_{b \in  \mathcal{N}_{\mathrm{BP}}(i)\setminus a}m_{f_b\to x_{i}}^{\mathrm{BP} }(x_{i})\prod_{c \in  \mathcal{N}_{\mathrm{MF}}(i)}m_{f_c\to x_{i}}^{\mathrm{MF}}(x_{i}),~\forall ai \in \mathcal{E}_{\mathrm{BP}},
    \\
    \label{eq:HtoMF}
    m_{f_a\to h_{l}}^{\mathrm{MF} }(h_{l}) &=& \mathrm{exp}\Big \{\left\langle\mathrm{ln}f_a \right\rangle_{b(\boldsymbol{x}_{\mathcal{N}_{\mathrm{BP}}(a)})\prod_{j\in \mathcal{N}_{\mathrm{MF}}(a)\setminus l} b(h_{j})} \Big \} ,~\forall al\in \mathcal{E}_{\mathrm{MF}},
    \\
    \label{eq:HBelief}
    b(\boldsymbol{x}_{\mathcal{N}_{\mathrm{BP}}(a)})&=& \mathrm{exp}\Big \{\left\langle\mathrm{ln}f_a \right\rangle_{\prod_{j\in  \mathcal{N}_{\mathrm{MF}}(a)}b(h_{j})} \Big \}\prod_{k\in \mathcal{N}_{\mathrm{BP}}(a)}n_{x_{k}\to f_a}(x_{k}),
    \\
    \label{eq:nMFHMP}
    n^{\mathrm{MF}}_{h_l\to f_a}(h_{l})&=&b(h_l) \propto \prod_{b \in  \mathcal{N}_{\mathrm{BP}}(l)}m_{f_b\to h_{l}}^{\mathrm{BP}}(h_{l})\prod_{c \in  \mathcal{N}_{\mathrm{MF}}(l)}m_{f_c\to h_{l}}^{\mathrm{MF}}(h_{l}), ~\forall al \in \mathcal{E}_{\mathrm{MF}},
\end{eqnarray} 
where $b(h_l)$ denotes the belief of the variable $h_l$. The factor nodes $\mathcal{A}_{\mathrm{Hybrid}}$ are able to connect variable nodes through different types of edges $\mathcal{E}_{\mathrm{BP}}$ and $\mathcal{E}_{\mathrm{MF}}$. In this way, HMP rule can implement two message calculation rules from the same factor node to the connected variable nodes by different edges. As shown in Fig. \ref{fig:HMP}, we assume that $f_a\in \mathcal{A}_{\mathrm{Hybrid}}$\footnote{Here, we simplify the expression $f_a \triangleq f_{x_1,\cdots,x_N,h_1,\cdots,h_L}.$}, variables $x_1,\cdots,x_N$ are connected to $f_a$ by BP edges; while variables $h_1,\cdots,h_L$ are connected to $f_a$ by MF edges. Then, the messages from factor node $f_a$ to $x_i, ai\in \mathcal{E}_{\mathrm{BP}}$ are calculated by \eqref{eq:HtoBP}, meanwhile the messages from the same factor node $f_a$ to $h_l, al\in \mathcal{E}_{\mathrm{MF}}$ are computed using \eqref{eq:HtoMF} and \eqref{eq:HBelief}. The opposite direction of the messages, i.e., from $x_i, ai\in \mathcal{E}_{\mathrm{BP}}$ and $h_l, al\in \mathcal{E}_{\mathrm{MF}}$ to $f_a$, are calculated by~\eqref{eq:nBPHMP} and~\eqref{eq:nMFHMP}, respectively.

We can analyze the relationships among the different rules from the perspective of the edge. i) If all the edges in a factor graph belong to $\mathcal{E}_{\mathrm{BP}}$, then the factor graph uses the BP rule only. ii) Instead, if all the edges in a factor graph belong to $\mathcal{E}_{\mathrm{MF}}$, then the factor graph uses the MF rule. iii) If a factor graph contains two types of factor nodes $\mathcal{A}_{\mathrm{BP}},\mathcal{A}_{\mathrm{MF}}$ and only one type of edge exists surrounding each type of factor node, then the factor graph uses the combined BP-MF rule. iv) If there are two types of edges surrounding a factor node in a factor graph, then the factor graph uses the HMP rule. Besides, if all the edges connected between a certain $\mathcal{A}_{\mathrm{Hybrid}}$ and its related variable nodes belong to $\mathcal{E}_{\mathrm{BP}}$, then the HMP rule at this factor node degenerates into the BP rule. Consequently, if all edges connected between a certain $\mathcal{A}_{\mathrm{Hybrid}}$ and its related variable nodes belong to $\mathcal{E}_{\mathrm{MF}}$, then, the HMP rule at this factor node degenerates into the MF rule. Compared with single and combined rules, HMP is more flexible to design message passing algorithms and therefore more suitable for solving complicated variational inference problems.

\section{System Model and Graph Representation}
\label{sec:model}
In this section, we present the system model of our inference problem and its factor graph representation. We consider a \revise{single-cell} downlink massive MIMO-OFDM system consisting of a single base station (BS) equipped with $N$ antennas, organized in a half-wavelength spacing uniform linear array (ULA), serving multiple single-antenna users. 
\revise{We assume that a resource grid of $K$ subcarriers and $T$ OFDM slots are available for communication in the scenario; it is further assumed that the channel coefficient of each subcarrier remains constant during the duration of $T$ slots. In the resource grid, $P < K$ subcarriers and $M < T$ slots are assigned to transmit pilot symbols, and used to perform the CE.}
We denote the downlink channel as $\boldsymbol{h}_{f}^{\left (p\right )}\in \mathbb{C}^{N\times 1}$, where the subscript $f$ represents the frequency domain of the vector, while the superscript $p$ indicates the $p$-th subcarrier, $ 1\le p\le P$. In the remainder of this article, we use the 3GPP spatial channel model (SCM) to generate the massive MIMO-OFDM channel~\cite{SCM}. 
To estimate the downlink channel, the BS sends $M$ training sequences $\boldsymbol{x}_{m}^{(p)}\in \mathbb{C}^{N\times 1}$, $1\le m\le M$ \revise{in the reserved time slots}. The matrix collecting all the training sequences is denoted as $\boldsymbol{X}^{\left (p\right)} = [ \boldsymbol{x}_{1}^{\left (p\right)},\cdots,\boldsymbol{x}_{M}^{(p)}]^{\mathrm{H}}\in \mathbb{C}^{M\times N}$. We assume that  $\boldsymbol{X}^{(p)}$ is a Partial DFT Random Permutation (PDFT-RP) pilot matrix~\cite{STCSMP}, which is proved to outperform than an i.i.d. Gaussian pilot matrix~\cite{TurboDFT}.
Thus, the received signal at users $\boldsymbol{y}_{f}^{\left (p\right)}$ can be written as
\begin{equation}
    \label{eq:receive}
    \boldsymbol{y}_{f}^{\left (p\right)} = \boldsymbol{X}^{\left (p\right)}\boldsymbol{h}_{f}^{\left (p\right)} + \boldsymbol{\omega}_{f}^{\left (p\right)}, \quad 1\le p\le P, 
\end{equation}
where $\boldsymbol{\omega}_{f}^{\left (p\right)}\sim\mathcal{CN}\left(\mathbf{0},\sigma^2 \boldsymbol{I} \right)$ denotes the additive white Gaussian noise (AWGN) with zero mean and variance $\sigma^2$. Following the approaches of~\cite{tse2005s, EBSLiuAn, Gao2016FDD}, we can transform the channel in the frequency domain $\boldsymbol{h}_{f}^{\left (p\right)}$ to the AF domain denoted by $\boldsymbol{h}_{a}^{\left (p\right)}$, as follows,
\begin{equation}
    \label{eq:Transform}
    \boldsymbol{h}_{a}^{\left (p\right)} = \boldsymbol{B}^{\mathrm{H}}\boldsymbol{h}_{f}^{\left (p\right)}, \quad 1\le p\le P, 
\end{equation}
where $\boldsymbol{B}$ is the unitary matrix representing the transformation of
the virtual angular domain at the BS side. Due to the BS equipped with ULA antennas, $\boldsymbol{B}$ is the discrete Fourier transform (DFT) matrix~\cite{tse2005s}. Substituting~\eqref{eq:Transform} into~\eqref{eq:receive}, and letting $\boldsymbol{A}^{(p)} = \boldsymbol{X}^{(p)}\boldsymbol{B}$, the received signal can be formulated as
\begin{equation}
    \label{eq:angulardomain}
    \boldsymbol{y}_{f}^{\left (p\right)} = \boldsymbol{A}^{\left (p\right)}\boldsymbol{h}_{a}^{\left (p\right)} + \boldsymbol{\omega}_{f}^{\left (p\right)}, \quad 1\le p\le P. 
\end{equation}
In this manner, we can focus on the estimation of the AF domain channel $\boldsymbol{h}_{a}^{\left (p\right)}$ based on the observed signal $\boldsymbol{y}_{f}^{\left (p\right)}$, $1\le p\le P$. Due to the channel sparsity as shown in the following subsection, a channel prior model is an essential aspect to solve the problem, such as~\cite{STCSMP,ChenSTCS,STCSFS}. However, these existing methods use the Bernoulli-Gaussian as the probability model, which can not accurately characterize the channel. Therefore, we analyze the channel data and propose a more appropriate probability model in the next subsection.

\begin{figure}
\centering
    \begin{subfigure}[t]{.49\textwidth}
        \centering
        \includegraphics[scale=0.55]{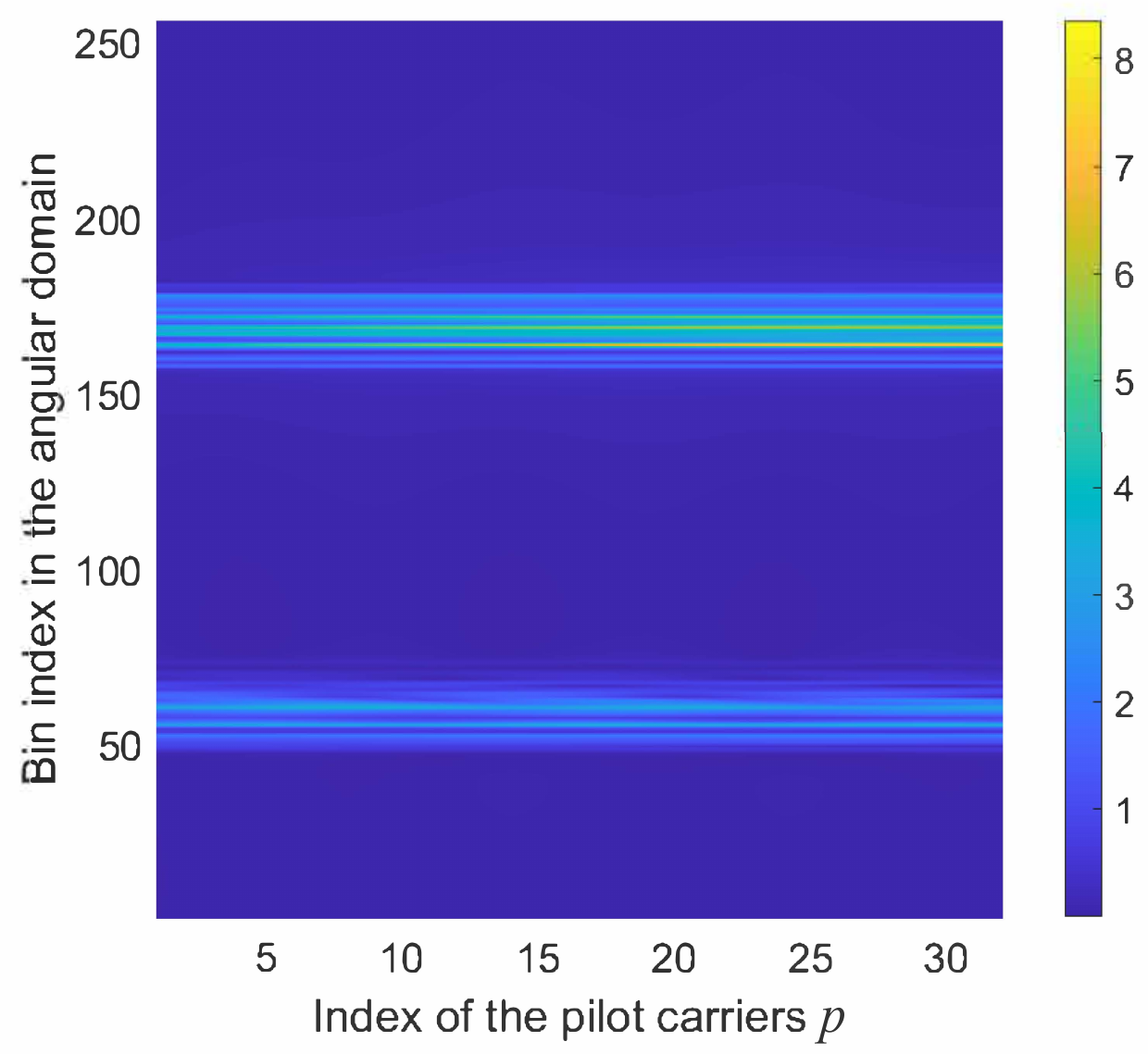}
        \caption{Channel gains in the angular-frequency domain.}\label{fig:SCM_all}
    \end{subfigure}
    \hfill
    \begin{subfigure}[t]{.49\textwidth}
        \centering
        \includegraphics[scale=0.55]{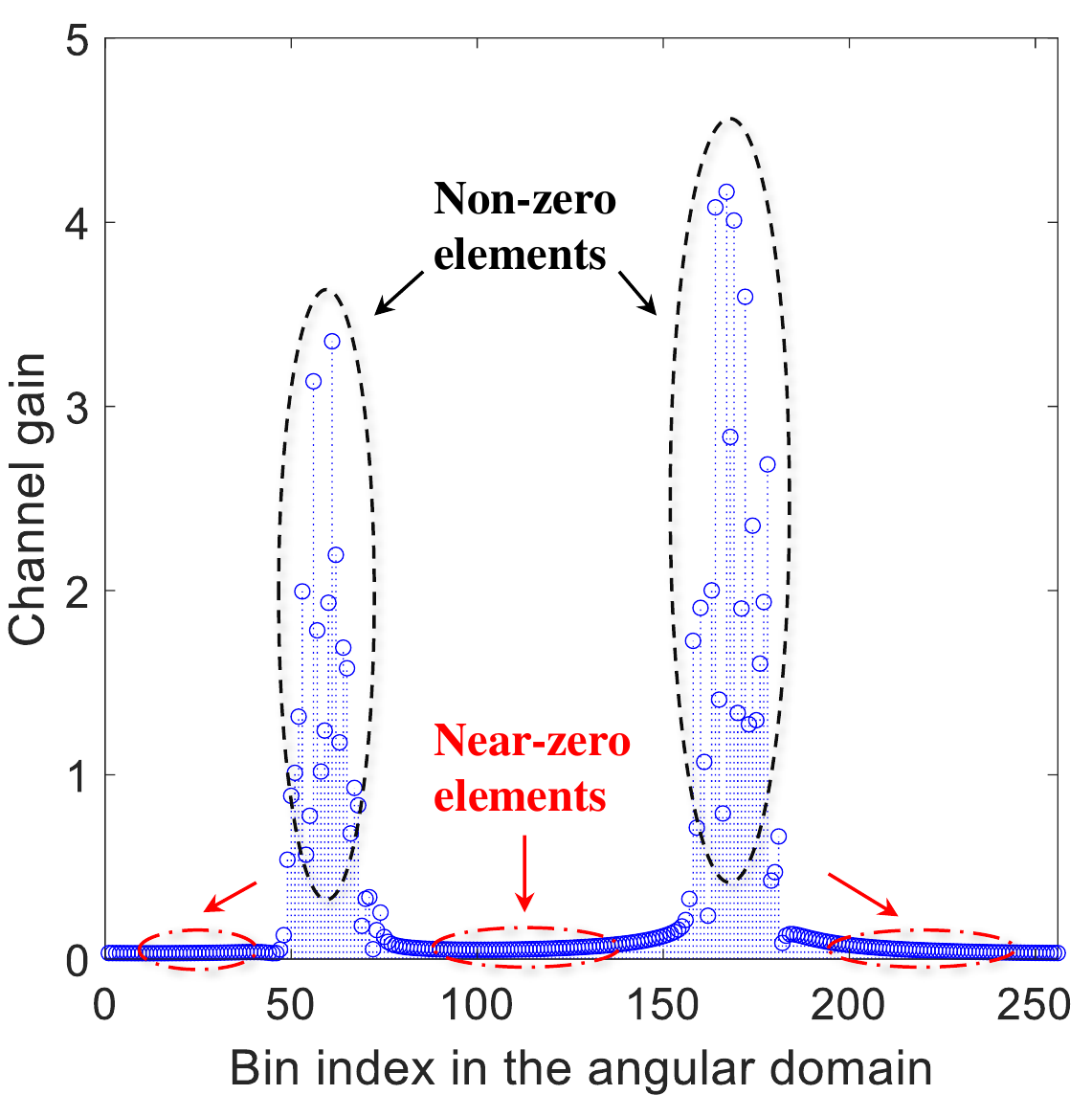}
        \caption{Channel gains of a particular subcarrier.}\label{fig:SCM_single}
    \end{subfigure}
\caption{A realization of the spatial channel model in~\cite{SCM}, with carrier frequency at 2 GHz, bandwidth 7.5 MHz and frequency interval 15 kHz. There are 512 subcarriers in total, and 32 of them are chosen as pilot subcarriers. The channel gains in subfigure (b) is $|\boldsymbol{h}_a^{(p)}|$.}
\label{fig:SCM}
\end{figure}

\subsection{Clustered-sparse Probability Model}
\label{sec:clusteredsparse}
The massive MIMO-OFDM channel exhibits clustered sparsity due to the scattering effect. Furthermore, the scatterers for different subchannels are quite similar~\cite{tse2005s}. Consequently, for a communication system with a much smaller bandwidth than the carrier frequency (as it is usually in the case of sub-6GHz communications), the subchannels $\{\boldsymbol{h}_{a}^{\left (p\right)}\}_{p=1}^{P}$ have a common support for sparsity~\cite{tse2005s}, i.e., 
\begin{equation}
    \label{eq:Common}
    \mathrm{supp}\{\boldsymbol{h}_{a}^{(1)}\} = \mathrm{supp}\{\boldsymbol{h}_{a}^{(2)}\} = \cdots =\mathrm{supp}\{\boldsymbol{h}_{a}^{(P)}\},
\end{equation}
where $\mathrm{supp}\{\boldsymbol{h}_{a}^{(p)}\}$ returns the positions of the non-zero entries of $\boldsymbol{h}_{a}^{(p)}$. 
In Fig.~\ref{fig:SCM}, we show a realization of the SCM urban macro scenario, generated by the parameters given in Table~\ref{tab:Parameter}. Fig.~\ref{fig:SCM}(a) shows that channel elements are sparse and clustered together. The positions of the non-zero elements (light-coloured data) among the different subcarriers exhibit common support, verifying~\eqref{eq:Common}. In addition, as shown in Fig.~\ref{fig:SCM}(b), the smaller values of the channel elements are not zero, but are close to zero. In the remainder of the article, we call these elements as ``\emph{near-zero}" elements. Besides, we find that the values of the non-zero elements vary significantly with respect to the antenna index, while the values of the near-zero elements are approximately constant.

Taking into consideration the above channel characteristics, we develop the TSGM-LVD probability model, i.e., at the same subcarrier, the variance differs among the non-zero elements while it is the same among the near-zero elements. The probability model can be written as
\begin{equation}
    \label{eq:Priormodel}
    p\big(h_{a,n}^{(p)}|s_n,v_{\mathrm{S}}^{(p)},v_{\mathrm{L},n}^{(p)}\big)=\delta \left(s_n\right ) ~\mathcal{CN}\big(h_{a,n}^{(p)};0,v_{\mathrm{S}}^{(p)-1}\big)+\delta(1-s_{n})~\mathcal{CN}\big(h_{a,n}^{(p)};0,v_{\mathrm{L},n}^{(p)-1}\big),
\end{equation}
where $\delta(\cdot )$ denotes the Dirac delta function, $s_{n}\in \left\{0,1\right\}$ is the hidden binary state indicating if the channel element is non-zero ($s_n = 1$) or near-zero ($s_n = 0$), $v_{\mathrm{L},n}^{(p)-1}$ denotes the variance of non-zero elements, and $v_{\mathrm{S}}^{(p)-1}$ denotes the variance of near-zero elements. 
%
Then, the clustering effect of the non-zero elements can be modeled by a Markov chain as follows
\begin{equation}
    \label{eq:Markovchain}
    p(\boldsymbol s)=p(s_{1})\prod_{n=2}^{N}p(s_{n}|s_{n-1}),
\end{equation}
with the transition and initial probabilities given by
\begin{equation}
    \label{eq:Transition}
    p(s_{n}|s_{n-1}) =\left\{\begin{matrix}(1-p_{10})^{(1-s_{n})}(p_{10})^{(s_{n})}, \quad s_{n-1}=0;
    \\
    (p_{01})^{(1-s_{n})}(1-p_{01})^{(s_{n})}, \quad s_{n-1}=1.\end{matrix}\right.
\end{equation}
and
\begin{equation}
    \label{eq:Initial}
    p(s_{1}) =(p_{10})^{ (s_1)}(1-p_{10})^{(1-s_1)}.
\end{equation}
The Markov chain can be characterized by parameters $p_{10}\triangleq \mathrm{Pr}(s_n=1|s_{n-1}=0) $ and $p_{01}\triangleq \mathrm{Pr}(s_n=0|s_{n-1}=1) $ thoroughly. Here, $p_{10}$ indicates the ``spacing'' between two clusters: on the one hand, a low value of $p_{10}$ means that the transition probability from state 0 to state 1 is relatively small, i.e., there is a large probability that more near-zero elements will be clustered together. On the other hand, $p_{01}$ reveal the size of the non-zero clusters: a low value of $p_{01}$ indicates a high probability that more non-zero elements will be clustered together. 

Prior works, such as~\cite{STCSFS, ChenSTCS}, initialize the probability in~\eqref{eq:Initial} using the the sparsity variable $\lambda \triangleq \mathrm{Pr}(s_{n}=1)=(1+p_{01}/p_{10})^{-1}$, which indicates the average ratio of the non-zero elements in $\boldsymbol{s}$.
When $N$ tends to infinity, the initialization $\mathrm{Pr}(s_1=1)=(1+p_{01}/p_{10})^{-1}$ is accurate~\cite{BICMOFDM}. However, in practice, when the BS has a limited number of antennas, initializing~\eqref{eq:Initial} as $\mathrm{Pr}(s_{1}=1)=p_{10}$ is more reasonable. 

Regarding the initialization of the other variables, we use a different approach instead of the EM algorithm, e.g.~\cite{ChenSTCS, STCSMP, STCSFS}. 
The hyperprior of the precision terms $v_{\mathrm{L},n}^{(p)}$ and $v_{\mathrm{S}}^{(p)}$ are assumed to be Gamma distributed\footnote{Note that, as in~\cite{tippingSBL}, we use the Gamma distribution for the parameter of precision, rather than for the variance~\cite{Carles}.}. In this way, the prior distribution of the precision terms are up to specific unknown parameters (see e.g.,~\cite{BPMFstretched}). The distributions result
\begin{equation}
    \label{eq:PrecisionGamma}
        p(v_{\mathrm{L},n}^{(p)})=\mathrm{Ga} (v_{\mathrm{L},n}^{(p)};\epsilon_n^{(p)},\eta_n^{(p)}),~~p(v_{\mathrm{S}}^{(p)}) =\mathrm{Ga} (v_{\mathrm{S}}^{(p)};\alpha^{(p)},\beta^{(p)}), 
\end{equation}
where we control the different initial values of $v_{\mathrm{L},n}^{(p)}$ and $v_{\mathrm{S}}^{(p)}$ by controlling the parameters $\epsilon_n^{(p)},\eta_n^{(p)}$ and $\alpha^{(p)},\beta^{(p)}$ of the Gamma distributions, respectively. 
Similarly, we also deem $p_{01}$ and $p_{10}$ as variables, generated by the Beta distributions as 
\begin{equation}
    \label{eq:ProbabilityBeta}
        p\left(p_{01}\right)=\mathrm{Beta} \left(p_{01};c,d\right),~~p\left(p_{10}\right) =\mathrm{Beta} \left(p_{10};e,f\right),
\end{equation}
where the parameters $c, d$ and $e, f$ determine the different initial values of $p_{01}$ and $p_{10}$.

\begin{figure}
    \begin{centering}
    \includegraphics[scale=0.8]{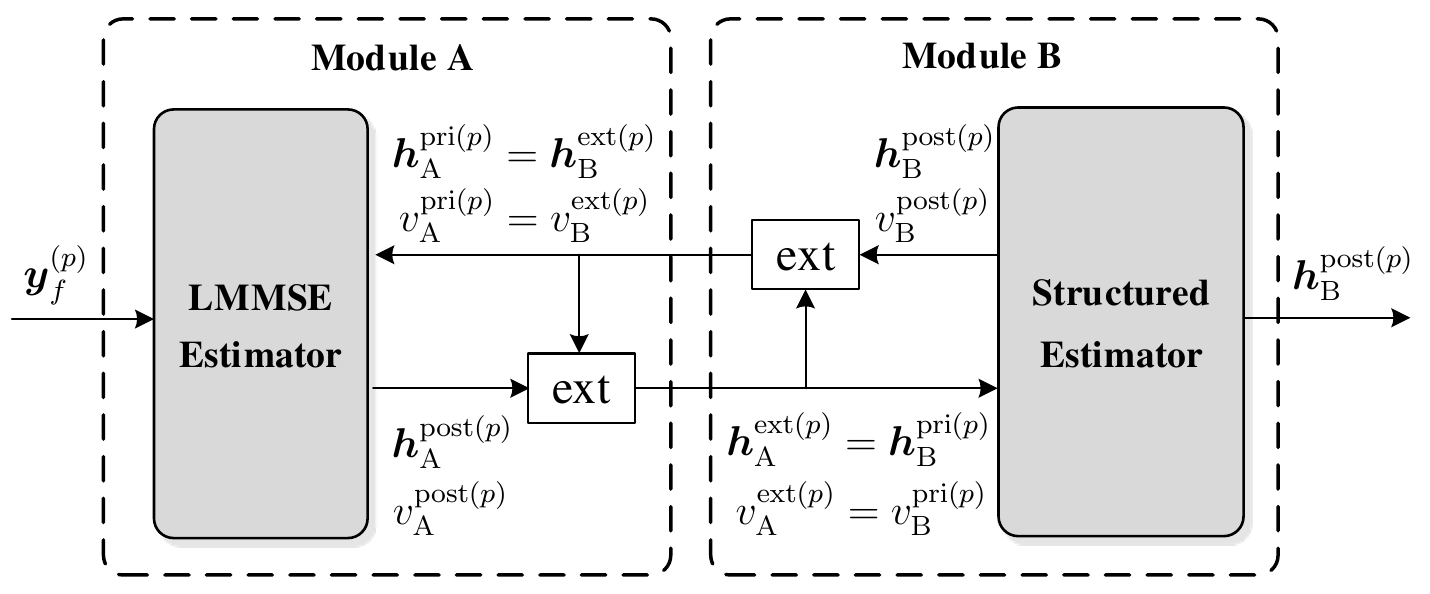}
    \caption{The block diagram of structured Turbo framework}
    \label{fig:STF}
    \end{centering}
\end{figure}

\subsection{Structured Turbo Framework}
\label{sec:STF}
To perform the CE, we use the structured Turbo framework (STF)~\cite{TurboDFT}, in conjunction with the channel probability model described in Section~\ref{sec:clusteredsparse}. 
The STF diagram consists in two modules~\cite{STCSMP}, as shown in Fig.~\ref{fig:STF}. 
\begin{itemize}
    \item \emph{Module A} comprises a linear minimum mean square error (LMMSE) estimator, which utilize the measurement $\boldsymbol{y}_f^{(p)}$ as well as prior message $\mathcal{CN} (\boldsymbol{h}_{a}^{(p)};\boldsymbol{h}_{\mathrm{A}}^{\mathrm{pri}(p)},v_{\mathrm{A}}^{\mathrm{pri}(p)}) $ from module B and output the \emph{posteriori} message $\mathcal{CN} (\boldsymbol{h}_{a}^{(p)};\boldsymbol{h}_{\mathrm{A}}^{\mathrm{post}(p)},v_{\mathrm{A}}^{\mathrm{post}(p)})$;
    
    \item \emph{Module B} includes a structured estimator, which can combine the channel probability model and prior message $\mathcal{CN} (\boldsymbol{h}_{a}^{(p)};\boldsymbol{h}_{\mathrm{B}}^{\mathrm{pri}(p)},v_{\mathrm{B}}^{\mathrm{pri}(p)})$  from module A to achieve better performance, and output $\boldsymbol{h}_{\mathrm{B}}^{\mathrm{post}(p)}$. 
\end{itemize}
The two models compute the extrinsic messages\footnote{The extrinsic messages are calculated by dividing two Gaussian PDFs, e.g., $v_{\mathrm{A}}^{\mathrm{ext}(p)-1}=v_{\mathrm{A}}^{\mathrm{post}(p)-1}-v_{\mathrm{A}}^{\mathrm{pri}(p)-1}, \boldsymbol{h}_{\mathrm{A}}^{\mathrm{ext}(p)}=v_{\mathrm{A}}^{\mathrm{ext}(p)}(\boldsymbol{h}_{\mathrm{A}}^{\mathrm{post}(p)}/v_{\mathrm{A}}^{\mathrm{post}(p)}-\boldsymbol{h}_{\mathrm{A}}^{\mathrm{pri}(p)}/v_{\mathrm{A}}^{\mathrm{pri}(p)})$.}
$\boldsymbol{h}_{\mathrm{A}}^{\mathrm{ext}(p)},v_{\mathrm{A}}^{\mathrm{ext}(p)}$ and $\boldsymbol{h}_{\mathrm{B}}^{\mathrm{ext}(p)},v_{\mathrm{B}}^{\mathrm{ext}(p)}$ iteratively until the algorithm converges. 

The STF utilized in this work is the same as~\cite{STCSFS, TurboDFT, ChenSTCS, STCSMP, DenoTSC}. The parameters used in module A and the evaluation of the extrinsic messages between the two modules are detailed in~\cite[Algorithm 1]{ChenSTCS}. 
\revise{}
The contributions of this article lie in the design of module B: the structured estimator uses the TSGM-LVD probability model to capture the channel sparsity more accurately as described in Section~\ref{sec:clusteredsparse}. Based on this, we construct the factor graph and design the MP algorithm for the realization of module B, as described in the following Sections.

\subsection{Probability Representation and Factor Graph}
\label{sec:factorgraph}
\begin{figure}
    \begin{centering}
        \includegraphics[scale=0.8]{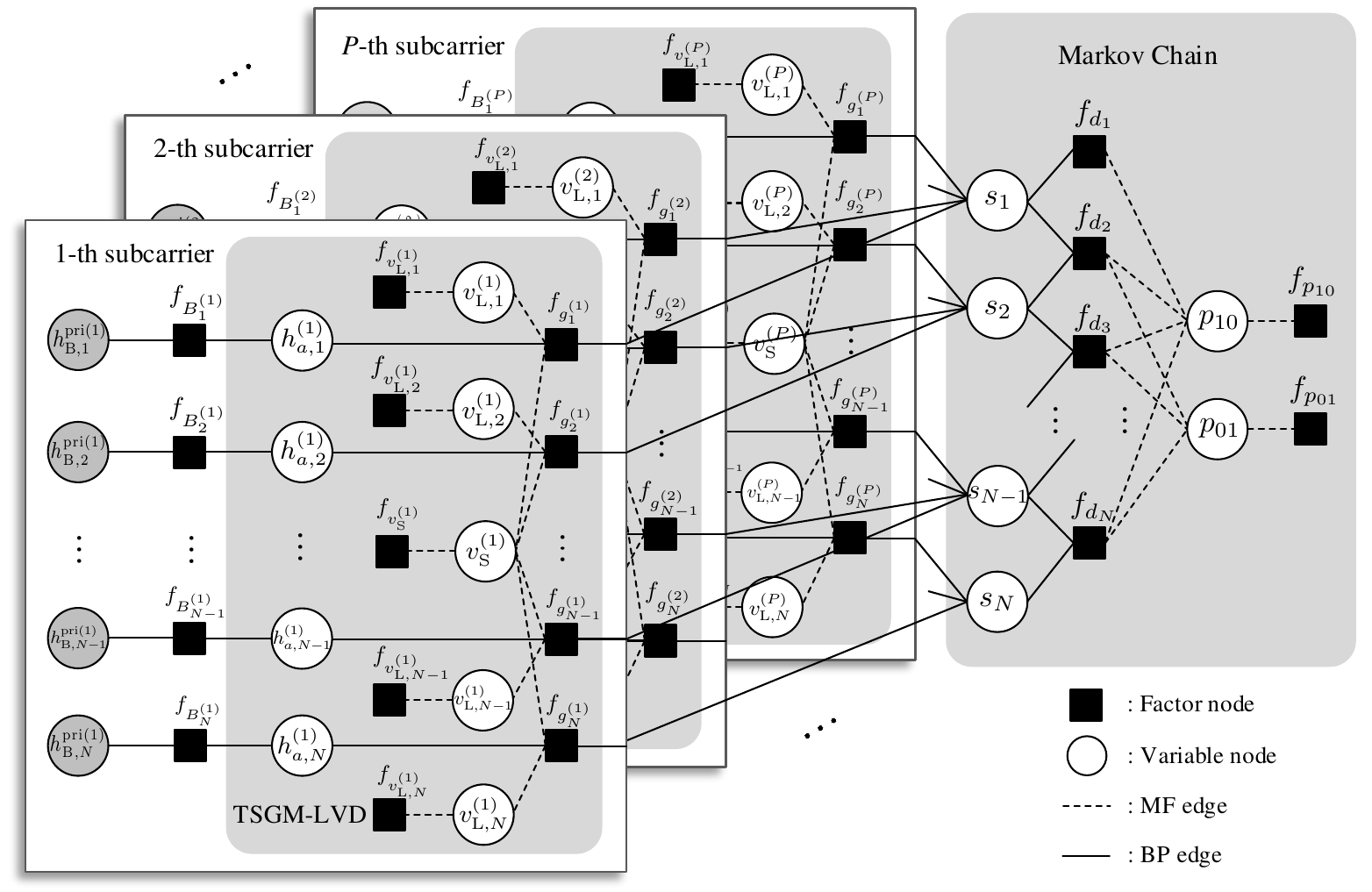}
        \caption{Factor graph representing the probability model of \eqref{eq:factorize}. The white circles, black squares and shaded gray circles represent the variable nodes,  factor nodes and the observations, respectively. Note that the observations are the extrinsic messages coming from module A.}
        \label{fig:systemfactor}
    \end{centering}
\end{figure}

In this subsection, we use Bayesian theory to model module B in the AF domain and represent it as a factor graph. A basic assumption is modeling $\boldsymbol h^{\mathrm{pri}(p)}_{\mathrm{B}}$ as an AWGN observation, i.e.,
\begin{equation}
    \label{eq:assumeB}
    \boldsymbol h^{\mathrm{pri}(p)}_{\mathrm{B}} = \boldsymbol h^{(p)}_{a}+\boldsymbol n^{(p)}_{a}, \quad 1 \le  p \le P,
\end{equation}
where $\boldsymbol n^{(p)}_{a}\sim\mathcal{CN} (\mathbf{0},v^{\mathrm{pri}(p)}_{\mathrm{B}}\boldsymbol{I}) $ is independent from $\boldsymbol h^{(p)}_{a}$. Similar assumption have been commonly applied in iterative signal recovery methods based on MP, such as\cite{TurboDFT, ChenSTCS, STCSFS, Donoho}. Under this assumption, and based on \eqref{eq:angulardomain}, we can factorize the joint PDF of all unknown random variables conditioned to the observation $\boldsymbol{Y}_f=[\boldsymbol y_{f}^{\left (1\right)},\cdots,\boldsymbol y_{f}^{\left (P\right)}]$ as
\begin{equation}
    \label{eq:factorize}
    \begin{split}
        p\left(\boldsymbol{H}_a,\boldsymbol{s},\boldsymbol{v}_{\mathrm{L}},\boldsymbol{v}_{\mathrm{S}},p_{10},p_{01} |\boldsymbol{Y}_f\right ) \propto \prod_{p=1}^{P}\prod_{n=1}^{N}f_{B_{n}^{(p)}}\big(h_{\mathrm{B},n}^{\mathrm{pri}(p)},h_{a,n}^{(p)}\big)f_{g_{n}^{(p)}}\big(h_{a,n}^{(p)},s_{n},v^{(p)}_{\mathrm{L},n},v^{(p)}_{\mathrm{S}}\big)
        \\
        \, f_{v_{\mathrm{L},n}^{(p)}}\big(v_{\mathrm{L},n}^{(p)}\big) \prod_{p=1}^{P}f_{v^{(p)}_{\mathrm{S}}}\big(v^{(p)}_{\mathrm{S} }\big) \prod_{n=2}^{N}f_{d_{n}}\left(s_{n},s_{n-1},p_{10},p_{01}\right)f_{d_{1}}\left(s_{1},p_{10}\right) f_{p_{01}}\left(p_{01}\right) f_{p_{10}}\left(p_{10}\right),
    \end{split}
\end{equation}
where $\boldsymbol{H}_a = [\boldsymbol{h}_a^{(1)},\cdots ,\boldsymbol{h}_a^{(P)}]$ is the collection of the AF domain channel vectors. The functions of factor nodes are listed in Table~\ref{tab:factorfunctions}, while the aforementioned factorization is illustrated in Fig.~\ref{fig:systemfactor}. Based on this factor graph, we will design the MP algorithm in the next Section.

\begin{table}[t]
	\centering
	\caption{Functions of the factor nodes in Fig. \ref{fig:systemfactor}.}
	\label{tab:factorfunctions}  
	\begin{tabular}{cc} 
		\hline\hline\noalign{\smallskip}	
		Factor Node  & Factor Function 
		\\
		\noalign{\smallskip}\hline\noalign{\smallskip}
		$f_{B_{n}^{(p)}}$ & 
		$p(h_{a,n}^{(p)}|h_{\mathrm{B},n}^{\mathrm{pri}(p)}) = 
        \mathcal{CN}(h_{a,n}^{(p)};h_{\mathrm{B},n}^{\mathrm{pri}(p)},v_{\mathrm{B}}^{\mathrm{pri}(p)})$  
        \\
		$f_{g_{n}^{(p)}}$ & $p(h_{a,n}^{(p)}|s_{n},v^{(p)}_{\mathrm{L},n},v_{\mathrm{S}}^{(p)}) =\delta(s_{n})
        \mathcal{CN}(h_{a,n}^{(p)};0,v^{(p)-1}_{\mathrm{S}})+\delta(1-s_{n})\mathcal{CN}(h_{a,n}^{(p)};0,v^{(p)-1}_{\mathrm{L},n})$ 
        \\
		$f_{v_{\mathrm{L},n}^{(p)}}$ & 
		$p(v_{\mathrm{L},n}^{(p)})=\mathrm{Ga}(v_{\mathrm{L},n}^{(p)};\epsilon ^{(p)}_{n},\eta^{(p)}_{n})$ 
		\\
		$f_{v_{\mathrm{S}}^{(p)}}$ & 
		$p(v_{\mathrm{S}}^{(p)})=\mathrm{Ga}(v_{\mathrm{S}}^{(p)};\alpha^{(p)},\beta^{(p)})$ 
		\\
		$f_{d_{n}}$ & 
		$p(s_{n},s_{n-1}|p_{10},p_{01})=\big[(1-p_{10})^{\delta(s_{n})}(p_{10})^{\delta(1-s_{n})}\big]^{\delta(s_{n-1})}\big[(p_{01})^{\delta(s_{n})}(1-p_{01})^{\delta(1-s_{n})}\big]^{\delta(1-s_{n-1})}$
		\\
		$f_{d_{1}}$ & 
		$p(s_{1}|p_{10})=(p_{10})^{\delta (1-s_1)}(1-p_{10})^{\delta (s_1)}$
		\\
		$f_{p_{01}}$ & 
		$p(p_{01})=\mathrm{Beta}(p_{01};c,d)$
		\\
		$f_{p_{10}}$ & 
		$p(p_{10})=\mathrm{Beta}(p_{10};e,f)$
		\\
		\noalign{\smallskip}\hline
	\end{tabular}
\end{table}

\section{Hybrid message passing channel estimation algorithm}
\label{sec:BPHF}
In this section, we propose the HMP-TSGM-LVD algorithm. Firstly, we divide all edges in the factor graph shown in Fig.~\ref{fig:systemfactor} into two categories based on the relationship between factors and variables, then calculate messages using the HMP rule, as described in Section \ref{sec:HMP}. Next, we present the message calculations in five parts, following the message delivery direction. Finally, we summarize the overall HMP-TSGM-LVD algorithm message scheduling.

\subsection{Calculation of Messages}
All the factors in Table~\ref{tab:factorfunctions} are collected in the set $\mathcal{A}_{\mathrm{Hybrid}}$; we group the edges as two disjoint subsets $\mathcal{E}_{\mathrm{BP}}$ and $\mathcal{E}_{\mathrm{MF}}$, as shown in Fig.~\ref{fig:systemfactor}. We use the solid lines to denote BP edges, dashed lines for MF edges. The message calculations on the BP edges employ~\eqref{eq:HtoBP} and~\eqref{eq:nBPHMP}, while~\eqref{eq:HtoMF},~\eqref{eq:HBelief} and~\eqref{eq:nMFHMP} are used to calculate the messages on the MF edges. 
We divide the messages calculation into five parts, according to the direction of the delivered message: right part, downward part, upward part\footnote{The downward and upward parts are based on the Markov chain formulation, and their complete name should be named Markov chain downward and Markov chain upward parts. For the sake of simplicity, the term Markov chain will be omitted in the following Sections.}, transition probability hyperparameters update part and left part. 

\subsubsection{Right part messages}
since the edge connect with $f_{B^{(p)}_{n}}$ and $h_{a,n}^{(p)}$ belong to $\mathcal{E}_{\mathrm{BP}}$, the extrinsic message passed from $f_{B^{(p)}_{n}}$ to $h_{a,n}^{(p)}$ can be expressed as
\begin{eqnarray}
    \label{eq:liklihood}
    m^{\mathrm{BP}}_{f_{B^{(p)}_{n}}\to h_{a,n}^{(p)}}\left(h_{a,n}^{(p)}\right)=\mathcal{CN}\left(h_{a,n}^{(p)};h_{\mathrm{B},n}^{\mathrm{pri}(p)},v_{\mathrm{B}}^{\mathrm{pri}(p)}\right), 
\end{eqnarray} 
where $h_{\mathrm{B},n}^{\mathrm{pri}(p)},v_{\mathrm{B}}^{\mathrm{pri}(p)}$ are based on AWGN observation, as described in~\eqref{eq:assumeB}.
Note that $n^{\mathrm{BP}}_{h_{a,n}^{(p)}\to f_{g^{(p)}_{n}}}(h_{a,n}^{(p)})$ is the same as $m^{\mathrm{BP}}_{f_{B^{(p)}_{n}}\to h_{a,n}^{(p)}}(h_{a,n}^{(p)})$. 
According to the assumption of Gamma distributed precision terms, the beliefs $b(v_{\mathrm{L},n}^{(p)})$ and $b(v_{\mathrm{S}}^{(p)})$ follow  $\mathrm{Ga}(v_{\mathrm{L},n}^{(p)};\hat{\epsilon}_n^{(p)},\hat{\eta}_n^{(p)} ) $ and $\mathrm{Ga}(v_{\mathrm{S}}^{(p)};\hat{\alpha}^{(p)},\hat{\beta}^{(p)}) 
$, whose parameters are updated in~\eqref{eq:epcUpdate} and~\eqref{eq:alphaUpdate}, respectively. Then, the message $m^{\mathrm{BP}}_{f_{g^{(p)}_{n}}\to s_n}(s_n)$ uses the HMP rule~\eqref{eq:HtoBP}, resulting in
\begin{equation} \label{eq:Sleft}
    \begin{aligned}
    m^{\mathrm{BP}}_{f_{g^{(p)}_{n}}\to s_n}(s_n) &= \int \mathrm{exp}\Big\{\big< \mathrm{ln}f_{g_n^{(p)}} \big>_{b(v_{\mathrm{S}}^{(p)})b(v_{\mathrm{L},n}^{(p)})}\Big\} n_{h_{a,n}^{(p)}\to f_{g_n^{(p)}}} (h_{a,n}^{(p)})\mathrm{d} h_{a,n}^{(p)} \\
    &= {\rm{\mathord{\buildrel{\lower3pt\hbox{$\scriptscriptstyle\rightharpoonup$}} 
    \over \pi } }}_n^{(p)}\delta (1-s_n)+
    (1-{\rm{\mathord{\buildrel{\lower3pt\hbox{$\scriptscriptstyle\rightharpoonup$}} 
    \over \pi } }}_n^{(p)})\delta (s_n), 
    \end{aligned}
\end{equation} 
where
\begin{eqnarray}
    \label{eq:phiright}
    {\rm{\mathord{\buildrel{\lower3pt\hbox{$\scriptscriptstyle\rightharpoonup$}} 
    \over \pi}}}_n^{(p)} \triangleq
    \frac{\frac{\mathrm{e}^{\psi(\hat{\epsilon}_n^{(p)})}}{\hat{\epsilon}_n^{(p)}} 
    \mathcal{CN}\left(h_{\mathrm{B},n}^{\mathrm{pri}(p)};0,v_{\mathrm{B}}^{\mathrm{pri}(p)}+\frac{\hat{\eta}_n^{(p)}}{\hat{\epsilon}_n^{(p)}} \right)}
    {\frac{\mathrm{e}^{\psi(\hat{\epsilon}_n^{(p)})}}{\hat{\epsilon}_n^{(p)}} 
    \mathcal{CN}\left(h_{\mathrm{B},n}^{\mathrm{pri}(p)};0,v_{\mathrm{B}}^{\mathrm{pri}(p)}+\frac{\hat{\eta}_n^{(p)}}{\hat{\epsilon}_n^{(p)}}\right)+
    \frac{\mathrm{e}^{\psi(\hat{\alpha}^{(p)})}}{\hat{\alpha}^{(p)}}\mathcal{CN}\left(h_{\mathrm{B},n}^{\mathrm{pri}(p)};0,v_{\mathrm{B}}^{\mathrm{pri}(p)}+\frac{\hat{\beta}^{(p)}}{\hat{\alpha}^{(p)}}\right)},
\end{eqnarray} 
and $\psi(x)\triangleq \mathrm{ln}x- \frac{1}{2x}$.


\subsubsection{Downward part messages}
According to the assumption of beliefs $b(p_{10})=\mathrm{Beta}(p_{10};\hat{e},\hat{f})$ and $b(p_{10})=\mathrm{Beta}(p_{01};\hat{c},\hat{d})$, whose parameters are updated later in \eqref{eq:updateef} and~\eqref{eq:updatecd}, the message from factor node $f_{d_1}$ to variable node $s_{1}$ employs HMP rule~\eqref{eq:HtoBP}. The message results
\begin{eqnarray}
    \label{eq:downMS1}
    m^{\mathrm{BP}}_{f_{d_1}\to s_1}(s_1)&=&\mathrm{exp}\big\{\left \langle \mathrm{ln}f_{d_1} \right \rangle _{b(p_{10})}\big\} 
    = \lambda ^\downarrow _1\delta (1-s_1)+(1-\lambda ^\downarrow _1)\delta (s_1),
\end{eqnarray} 
where 
\begin{eqnarray}
    \label{eq:lamdadownMS1}
    \lambda ^\downarrow _1\triangleq\frac{\mathrm{exp}\left\{\left \langle\mathrm{ln}p_{10}\right \rangle\right\}}{\mathrm{exp}\left\{\left \langle\mathrm{ln}p_{10}\right \rangle\right\}+\mathrm{exp}\left\{\left \langle\mathrm{ln}(1-p_{10})\right \rangle\right\}},
\end{eqnarray} 
and
\begin{equation}
    \label{eq:lnBeta}
    \left \langle\mathrm{ln}p_{10}\right \rangle=\psi (\hat{e})-\psi (\hat{e}+\hat{f}),~~\left \langle\mathrm{ln}(1-p_{10})\right \rangle=\psi (\hat{f})-\psi (\hat{e}+\hat{f}).
\end{equation}
Then, the message from variable node $s_{n}$ to factor nodes $f_{d_{n+1}}$ uses \eqref{eq:nBPHMP}, obtaining
\begin{eqnarray}
    \label{eq:downNSn}
    n^{\mathrm{BP}}_{s_n\to f_{d_{n+1}}}(s_n)=\lambda ^\Downarrow  _n\delta (1-s_n)+(1-\lambda ^\Downarrow  _n)\delta (s_n) , ~~~ 1 \le n \le N-1,
\end{eqnarray} 
where
\begin{equation} 
\label{eq:lamdadownNSn}
\begin{aligned}
    \lambda^\Downarrow _n \triangleq\frac{\lambda ^\downarrow _n\prod_{p=1}^{P}{\rm{\mathord{\buildrel{\lower3pt\hbox{$\scriptscriptstyle\rightharpoonup$}} \over \pi}}}_n^{(p)} }
    {\lambda ^\downarrow _n\prod_{p=1}^{P}{\rm{\mathord{\buildrel{\lower3pt\hbox{$\scriptscriptstyle\rightharpoonup$}} \over \pi}}}_n^{(p)}  +
    (1-\lambda ^\downarrow _n)\prod_{p=1}^{P}(1-{\rm{\mathord{\buildrel{\lower3pt\hbox{$\scriptscriptstyle\rightharpoonup$}} \over \pi}}}_n^{(p)})},
\end{aligned}
\end{equation} 
Similar to (\ref{eq:downMS1}), the message $m^{\mathrm{BP}}_{f_{d_n}\to s_n}(s_n)$ is obtained by the HMP rule~\eqref{eq:HtoBP} as
\begin{equation} 
\label{eq:downMSn}
\begin{aligned}
    m^{\mathrm{BP}}_{f_{d_n}\to s_n}(s_n) &=\int \mathrm{exp}\big\{\left \langle \mathrm{ln}f_{d_n} \right \rangle _{b(p_{10})b(p_{01})}\big\} n_{s_{n-1}\to f_{d_n}}(s_{n-1})\mathrm{d} s_{n-1}
    \\
    &=\lambda ^\downarrow _n\delta (1-s_n)+(1-\lambda ^\downarrow _n)\delta (s_n),~~~2\le n \le N-1,
\end{aligned}
\end{equation}
where 
\begin{eqnarray}
    \label{eq:lamdadownMSn}
    \lambda ^\downarrow _n \triangleq\frac{\lambda ^\Downarrow_{n-1} \mathrm{Con1} +
     (1-\lambda ^\Downarrow_{n-1})
    \mathrm{Con2}}{\lambda ^\Downarrow_{n-1} (\mathrm{Con1}+\mathrm{Con4})+(1-\lambda ^\Downarrow_{n-1})(\mathrm{Con2}+\mathrm{Con3})},
\end{eqnarray} 
and
\begin{equation}
    \label{eq:constant}
    \begin{split}
        \mathrm{Con1} = \mathrm{exp}\left\{\left \langle \mathrm{ln}(1-p_{01})\right \rangle \right\} ;~~\mathrm{Con2} = \mathrm{exp}\left\{\left \langle\mathrm{ln}p_{10}\right \rangle\right\};  
        \\
        \mathrm{Con3}=\mathrm{exp}\left\{\left \langle\mathrm{ln}(1-p_{10})\right \rangle\right\};~~\mathrm{Con4} = \mathrm{exp}\left\{\left \langle\mathrm{ln}p_{01}\right \rangle\right\};
        \\
         \left \langle\mathrm{ln}p_{01}\right \rangle=\psi (\hat{c})-\psi (\hat{c}+\hat{d}),~~\left \langle\mathrm{ln}(1-p_{01})\right \rangle=\psi (\hat{d})-\psi (\hat{c}+\hat{d}).
    \end{split}
\end{equation}

\subsubsection{Upward part messages}
First, we initialize $\lambda ^\uparrow _N = 1/2$. The message $n^{\mathrm{BP}}_{s_n\to f_{d_{n}}}(s_n)$ from $s_{n}$ to $f_{d_{n}}$ results
\begin{eqnarray}
    \label{eq:upNSn}
    n^{\mathrm{BP}}_{s_n\to f_{d_{n}}}(s_n)=\lambda^\Uparrow_n\delta(1-s_n)+(1-\lambda^\Uparrow_n)\delta(s_n) ,~~~1\le n \le N,
\end{eqnarray} 
where
\begin{eqnarray}
    \label{eq:lamdaupNSn}
    \lambda^\Uparrow_n\triangleq\frac{\lambda ^\uparrow _n\prod_{p=1}^{P}{\rm{\mathord{\buildrel{\lower3pt\hbox{$\scriptscriptstyle\rightharpoonup$}} \over \pi}}}_n^{(p)} }
    {\lambda ^\uparrow _n\prod_{p=1}^{P}{\rm{\mathord{\buildrel{\lower3pt\hbox{$\scriptscriptstyle\rightharpoonup$}} \over \pi}}}_n^{(p)}+
    (1-\lambda ^\uparrow _n)\prod_{p=1}^{P}(1-{\rm{\mathord{\buildrel{\lower3pt\hbox{$\scriptscriptstyle\rightharpoonup$}} \over \pi}}}_n^{(p)})}.
\end{eqnarray} 
Then, we can compute the message from $f_{d_{n+1}}$ to $s_{n}$ by the HMP rule~\eqref{eq:HtoBP}, as follows
\begin{equation}\label{eq:upMSn}
\begin{aligned}
    m^{\mathrm{BP}}_{f_{d_{n+1}}\to s_n}(s_n) = 
    \lambda ^\uparrow _n\delta (1-s_n)+(1-\lambda ^\uparrow _n)\delta (s_n),~~~1\le n \le N-1,
\end{aligned}
\end{equation}
where 
\begin{eqnarray}
    \label{eq:lamdaupMSn}
    \lambda ^\uparrow _n\triangleq\frac{\lambda^\Uparrow_{n+1}\mathrm{Con1}+(1-\lambda^\Uparrow_{n+1}) \mathrm{Con4}}
    {\lambda ^\Uparrow_{n+1} (\mathrm{Con1}+\mathrm{Con2})+(1-\lambda^\Uparrow_{n+1})(\mathrm{Con3}+\mathrm{Con4})},
\end{eqnarray} 
where $\mathrm{Con1},\mathrm{Con2},\mathrm{Con3},\mathrm{Con4} $ are defined in~\eqref{eq:constant}.

\subsubsection{Transition probability hyperparameters update part messages}
The belief $b(s_1)$ and the combined belief $b(s_n,s_{n-1}),2\le n \le N$ are calculated by~\eqref{eq:HBelief}, resulting in
\begin{equation}
    \label{eq:beliefS1}
    b(s_1)=B_{s_1}\delta (1-s_1)+(1-B_{s_1})\delta (s_1), 
\end{equation}
\begin{equation}
    \label{eq:combelief}
    \begin{split}
        b(s_n,s_{n-1})=\frac{1}{\rho _{n,n-1}} \left[B_{n,n-1}^{00}\delta (s_n)\delta (s_{n-1})+B_{n,n-1}^{01}\delta (s_n)\delta (1-s_{n-1})+\right. 
        \\
        \left. B_{n,n-1}^{10}\delta (1-s_n)\delta (s_{n-1})+B_{n,n-1}^{11}\delta (1-s_n)\delta (1-s_{n-1}) \right],
    \end{split}
\end{equation}
where the normalized factors are
\begin{equation}
    \label{eq:Bs1}
    B_{s_1}\triangleq\frac{\lambda ^\uparrow_1\lambda ^\downarrow_1 {\textstyle \prod_{p=1}^{P}}{\rm{\mathord{\buildrel{\lower3pt\hbox{$\scriptscriptstyle\rightharpoonup$}} 
    \over \pi}}}_1^{(p)}}{\lambda ^\uparrow_1\lambda ^\downarrow_1 {\textstyle \prod_{p=1}^{P}}{\rm{\mathord{\buildrel{\lower3pt\hbox{$\scriptscriptstyle\rightharpoonup$}} 
    \over \pi}}}_1^{(p)}+(1-\lambda ^\uparrow_1)(1-\lambda ^\downarrow_1 ){\textstyle \prod_{p=1}^{P}}(1-{\rm{\mathord{\buildrel{\lower3pt\hbox{$\scriptscriptstyle\rightharpoonup$}} 
    \over \pi}}}_1^{(p)})} , 
\end{equation}
\begin{equation}
    \label{eq:Bcomsn}
    \begin{split}
        B_{n,n-1}^{00}&=(1-\lambda^\Uparrow_n)(1-\lambda^\Downarrow_{n-1})\mathrm{exp}\left\{\left \langle \mathrm{ln}(1-p_{10})\right \rangle \right\},
        \\
        B_{n,n-1}^{01}&=(1-\lambda^\Uparrow_n)\lambda^\Downarrow_{n-1}\mathrm{exp}\left\{\left \langle \mathrm{ln}p_{01}\right \rangle \right\},
         \\
         B_{n,n-1}^{10}&=\lambda^\Uparrow_n(1-\lambda^\Downarrow_{n-1})\mathrm{exp}\left\{\left \langle \mathrm{ln}p_{10}\right \rangle \right\},
         \\
        B_{n,n-1}^{11}&=\lambda^\Uparrow_n\lambda^\Downarrow_{n-1}\mathrm{exp}\left\{\left \langle \mathrm{ln}(1-p_{01})\right \rangle \right\},
        \\
        \rho _{n,n-1}&=B_{n,n-1}^{00}+B_{n,n-1}^{01}+B_{n,n-1}^{10}+B_{n,n-1}^{11}.
    \end{split}
\end{equation}
Then, we can apply HMP rule~\eqref{eq:HtoMF} to calculate the messages from $f_{d_{1}}$ to $p_{10}$ and $f_{d_{n}}$ to $p_{10},p_{01}$, $2\le n \le N$ as
\begin{equation}
    \label{eq:fd1to10}
        m_{f_{d_1}\to p_{10}}^{\mathrm{MF} }(p_{10}) =\mathrm{exp}\big\{\left \langle \mathrm{ln}f_{d_{1}} \right \rangle _{b(s_1)}\big\} \propto \mathrm{Beta}(p_{10};B_{s_1}+1,2-B_{s_1}),
\end{equation}
\begin{equation}
    \label{eq:fdnto10}
         m_{f_{d_n}\to p_{10}}^{\mathrm{MF}}(p_{10}) = \mathrm{exp}\big\{\left \langle \mathrm{ln}f_{d_{n}} \right \rangle _{b(s_n,s_{n-1})b(p_{01})}\big\} \propto \mathrm{Beta}\Big(p_{10};\frac{B_{n,n-1}^{10}}{\rho_{n,n-1}} +1,\frac{B_{n,n-1}^{00}}{\rho_{n,n-1}} +1\Big),
\end{equation}
\begin{equation}
    \label{eq:fdnto01}
        m_{f_{d_n}\to p_{01}}^{\mathrm {MF}}(p_{01}) =\mathrm{exp}\big\{\left \langle \mathrm{ln}f_{d_{n}} \right \rangle _{b(s_n,s_{n-1})b(p_{10})}\big\} 
        \propto \mathrm{Beta}\Big(p_{01};\frac{B_{n,n-1}^{01}}{\rho_{n,n-1}}+1,\frac{B_{n,n-1}^{11}}{\rho_{n,n-1}} +1\Big).
\end{equation}
Given $m^{\mathrm{MF}}_{f_{p_{10}}\to p_{10}}(p_{10})=\mathrm{Beta} (p_{10};e,f)$ and $m^{\mathrm{MF}}_{f_{p_{01}}\to p_{01}}(p_{01})=\mathrm{Beta} (p_{01};c,d)$, the beliefs $b(p_{10})$ and $b(p_{01})$ of variables $p_{10}$ and $p_{01}$ read
\begin{equation}
    \label{eq:beliefp10}
        b(p_{10})\propto \mathrm{Beta}(p_{10};\hat{e},\hat{f}),~~b(p_{01})\propto \mathrm{Beta}(p_{01};\hat{c},\hat{d}),
\end{equation}
whose parameters are updated through
\begin{align}
    \label{eq:updateef}
    \hat{e}&=B_{s_1}+e+\sum_{n=2}^{N} \frac{B_{n,n-1}^{10}}{\rho_{n,n-1}};\quad \hat{f}=1-B_{s_1}+f+\sum_{n=2}^{N}\frac{B_{n,n-1}^{00}}{\rho_{n,n-1}},
    \\
    \label{eq:updatecd}
    \hat{c}&=c+\sum_{n=2}^{N} \frac{B_{n,n-1}^{01}}{\rho_{n,n-1}};~~~~~~~~~~\hat{d}=d+\sum_{n=2}^{N}\frac{B_{n,n-1}^{11}}{\rho_{n,n-1}}.
\end{align}

\subsubsection{Left part messages}
We update downward and upward messages in \emph{Part~2} and \emph{Part~3} again using the beliefs $b(p_{10})$ and $b(p_{01})$ updated through \eqref{eq:updateef} and~\eqref{eq:updatecd}, to increase the estimation performance. Then, the message going out of the Markov chain $n^{\mathrm{BP}}_{s_n\to f_{g_n^{(p)}}}(s_n)$ from $s_n$ to $f_{g_n^{(p)}}$ is updated by 
\begin{equation}
    \label{eq:Sright}
        n^{\mathrm{BP}}_{s_n\to f_{g_n^{(p)}}}(s_n)={\rm{\mathord{\buildrel{\lower3pt\hbox{$\scriptscriptstyle\leftharpoonup$}}\over\pi}}}_n^{(p)}\delta(1-s_n)+(1-{\rm{\mathord{\buildrel{\lower3pt\hbox{$\scriptscriptstyle\leftharpoonup$}}\over\pi}}}_n^{(p)})\delta(s_n),~~~1\le n\le N,1\le p\le P, 
\end{equation}
where
\begin{eqnarray}
    \label{eq:phileft}
    {\rm{\mathord{\buildrel{\lower3pt\hbox{$\scriptscriptstyle\leftharpoonup$}} 
    \over \pi}}}_n^{(p)} \triangleq\frac{\lambda^\uparrow_n\lambda^\downarrow_n\prod_{{p}'\ne p}^{P}{\rm{\mathord{\buildrel{\lower3pt\hbox{$\scriptscriptstyle\rightharpoonup$}}\over \pi}}}_n^{({p}')}}
    {\lambda^\uparrow_n\lambda^\downarrow_n\prod_{{p}'\ne p}^{P}{\rm{\mathord{\buildrel{\lower3pt\hbox{$\scriptscriptstyle\rightharpoonup$}}\over \pi}}}_n^{({p}')}+
    (1-\lambda^\uparrow_n)(1-\lambda^\downarrow_n)\prod_{{p}'\ne p}^{P}(1-{\rm{\mathord{\buildrel{\lower3pt\hbox{$\scriptscriptstyle\rightharpoonup$}}\over \pi}}}_n^{({p}')})}.
\end{eqnarray} 
In order to use HMP rule to update the variances of the TSGM-LVD prior model, we first calculate the combined belief $b(h_{a,n}^{(p)},s_n)$ using \eqref{eq:HBelief} as
\begin{equation}
    \label{eq:beliefHS}
    \begin{aligned}
    b(h_{a,n}^{(p)},s_n)&=B_{h_n,s_n}^{(p)}\delta (1-s_n) \mathcal{CN} (h_{a,n}^{(p)}, \hat{\mu}_{\mathrm{L},n}^{(p)}, \hat{\varsigma}_{\mathrm{L},n}^{(p)} ) +(1-B_{h_n,s_n}^{(p)}) \delta (s_n)\mathcal{CN}(h_{a,n}^{(p)}, \hat{\mu}_{\mathrm{S},n}^{(p)}, \hat{\varsigma}_{\mathrm{S}}^{(p)}),
    \end{aligned}
\end{equation} 
where
\begin{equation}
    \label{eq:HSconstant}
    B_{h_n,s_n}^{(p)}\triangleq  \frac{{\rm{\mathord{\buildrel{\lower3pt\hbox{$\scriptscriptstyle\rightharpoonup$}} \over \pi}}}_n^{(p)}
    {\rm{\mathord{\buildrel{\lower3pt\hbox{$\scriptscriptstyle\leftharpoonup$}} \over \pi}}}_n^{(p)}}
    {{\rm{\mathord{\buildrel{\lower3pt\hbox{$\scriptscriptstyle\rightharpoonup$}} \over \pi}}}_n^{(p)}
    {\rm{\mathord{\buildrel{\lower3pt\hbox{$\scriptscriptstyle\leftharpoonup$}} \over \pi}}}_n^{(p)}+
    (1-{\rm{\mathord{\buildrel{\lower3pt\hbox{$\scriptscriptstyle\rightharpoonup$}} \over \pi}}}_n^{(p)})
    (1-{\rm{\mathord{\buildrel{\lower3pt\hbox{$\scriptscriptstyle\leftharpoonup$}} \over \pi}}}_n^{(p)})},  \\
\end{equation}
\begin{align}
    \label{eq:Largemean}
    \hat{\varsigma}_{\mathrm{L},n}^{(p)} &= (v_{\mathrm{B}}^{\mathrm{pri}(p)-1}+\hat{\epsilon}^{(p)}_n/\hat{\eta}^{(p)}_n )^{-1}, \quad
    \hat{\mu}_{\mathrm{L},n}^{(p)}=\hat{\varsigma}_{\mathrm{L},n}^{(p)}h_{\mathrm{B},n}^{\mathrm{pri}(p)}/v_{\mathrm{B}}^{\mathrm{pri}(p)} ; \\
    \label{eq:Smallmean}
    \hat{\varsigma}_{\mathrm{S}}^{(p)} &= (v_{\mathrm{B}}^{\mathrm{pri}(p)-1}+\hat{\alpha}^{(p)}/\hat{\beta}^{(p)})^{-1}, \quad
    \hat{\mu}_{\mathrm{S},n}^{(p)}=\hat{\varsigma}_{\mathrm{S}}^{(p)}h_{\mathrm{B},n}^{\mathrm{pri}(p)}/v_{\mathrm{B}}^{\mathrm{pri}(p)}.
\end{align}
Then, the messages from $f_{g_n^{(p)}}$ to variable nodes $v_{\mathrm{L},n}^{(p)}$ and $v_{\mathrm{S}}^{(p)}$ can be computed with the HMP rule \eqref{eq:HtoMF} as
\begin{equation}
    \label{eq:fgtovL}
    \begin{split}
    m^{\mathrm{MF}}_{f_{g_n^{(p)}}\to v_{\mathrm{L},n}^{(p)}}\big(v_{\mathrm{L},n}^{(p)}\big)&=\mathrm{exp}\Big\{\big< \mathrm{ln}f_{g_{n}^{(p)}} \big> _{b(h_{a,n}^{(p)},s_n)b(v_{\mathrm{S}}^{(p)})}\Big\}
    \\
    &\propto \mathrm{Ga}(v_{\mathrm{L},n}^{(p)};1+B_{h_n,s_n},B_{h_n,s_n}(|\hat{\mu}_{\mathrm{L},n}^{(p)}|^2+\hat{\varsigma }_{\mathrm{L},n}^{(p)})),
    \end{split}
\end{equation}
\begin{equation}
    \label{eq:fgtovS}
    \begin{split}
    m^{\mathrm{MF}}_{f_{g_n^{(p)}}\to v_{\mathrm{S}}^{(p)}}\big(v_{\mathrm{S}}^{(p)}\big)&=\mathrm{exp}\Big\{\big<\mathrm{ln}f_{g_{n}^{(p)}}\big>_{b(h_{a,n}^{(p)},s_n)b(v_{\mathrm{L},n}^{(p)})}\Big\}
    \\
    &\propto \mathrm{Ga}(v_{\mathrm{S}}^{(p)};2-B_{h_n,s_n},
    (1-B_{h_n,s_n})(|\hat{\mu}_{\mathrm{S},n}^{(p)}|^2+\hat{\varsigma }_{\mathrm{S}}^{(p)})).
    \end{split}
\end{equation}
Given the prior distributions of $v_{\mathrm{L},n}^{(p)}$ and $v_{\mathrm{S}}^{(p)}$, i.e., $m^{\mathrm{MF}}_{f_{v_{\mathrm{L},n}^{(p)}}\to v_{\mathrm{L},n}^{(p)}}(v_{\mathrm{L},n}^{(p)})=\mathrm{Ga} (v_{\mathrm{L},n}^{(p)};\epsilon_n^{(p)},\eta_n^{(p)})$ and $m^{\mathrm{MF}}_{f_{v_{\mathrm{S}}^{(p)}}\to v_{\mathrm{S}}^{(p)}}(v_{\mathrm{S}}^{(p)})=\mathrm{Ga} (v_{\mathrm{S}}^{(p)};\alpha^{(p)},\beta^{(p)})$, we can update the beliefs $b(v_{\mathrm{L},n}^{(p)})$ and $b(v_{\mathrm{S}}^{(p)})$ as
\begin{equation}
    \label{eq:beliesvL}
        b(v_{\mathrm{L},n}^{(p)}) =\mathrm{Ga}(v_{\mathrm{L},n}^{(p)};\hat{\epsilon}^{(p)}_n,\hat{\eta}^{(p)}_n),~~b(v_{\mathrm{S}}^{(p)}) =\mathrm{Ga}(v_{\mathrm{S}}^{(p)};\hat{\alpha}^{(p)},\hat{\beta}^{(p)}),
\end{equation}
whose parameters are updated through
\begin{align}
    \label{eq:epcUpdate}
    \hat{\epsilon}^{(p)}_n&=\epsilon^{(p)}_n+B_{h_n,s_n},~~~~~~~~~~~~~~~
    \hat{\eta}^{(p)}_n=\eta^{(p)}_n+B_{h_n,s_n}(|\hat{\mu}_{\mathrm{L},n}^{(p)}|^2+\hat{\varsigma }_{\mathrm{L},n}^{(p)});  
    \\
    \label{eq:alphaUpdate}
    \hat{\alpha}^{(p)}&=\alpha^{(p)}+\sum_{n=1}^{N}(1-B_{h_n,s_n}), \quad
    \hat{\beta}^{(p)}=\beta^{(p)}+\sum_{n=1}^{N}(1-B_{h_n,s_n})(|\hat{\mu}_{\mathrm{S},n}^{(p)}|^2+\hat{\varsigma }_{\mathrm{S}}^{(p)}).
\end{align}
With the parameters updated in ~\eqref{eq:epcUpdate} and~\eqref{eq:alphaUpdate}, we can compute the messages $m_{f_{g^{(p)}_{n}}\to h_{a,n}^{(p)}}^{\mathrm{BP}}(h_{a,n}^{(p)})$ from $f_{g_n^{(p)}}$ to $h_{a,n}^{(p)}$ by HMP rule \eqref{eq:HtoBP} as
\begin{equation}
    \label{eq:fgtoha}
    \begin{split}
        m_{f_{g^{(p)}_{n}}\to h_{a,n}^{(p)}}^{\mathrm{BP}}(h_{a,n}^{(p)})&=\int \mathrm{exp}\Big\{\big< \mathrm{ln}f_{g_{n}^{(p)}}\big> _{b(v_{\mathrm{L},n}^{(p)})b(v_{\mathrm{S}}^{(p)})}\Big\}n_{s_n\to f_{g_{n}^{(p)}}}(s_n)\mathrm{d} s_n
        \\
        &={\rm{\mathord{\buildrel{\lower3pt\hbox{$\scriptscriptstyle\leftharpoonup$}} \over \pi } }}_n^{(p)}\frac{\mathrm{e}^{\psi(\hat{\epsilon}_n^{(p)})}}{\hat{\epsilon}_n^{(p)}} \mathcal{CN}\Big(h_{a,n}^{(p)};0,\frac{\hat{\eta}_n^{(p)}}{\hat{\epsilon}_n^{(p)}} \Big)+(1-{\rm{\mathord{\buildrel{\lower3pt\hbox{$\scriptscriptstyle\leftharpoonup$}} \over \pi } }}_n^{(p)})\frac{\mathrm{e}^{\psi(\hat{\alpha}^{(p)})}}{\hat{\alpha}^{(p)}}\mathcal{CN}\Big(h_{a,n}^{(p)};0,\frac{\hat{\beta}^{(p)}}{\hat{\alpha}^{(p)}}\Big),
 \end{split}
\end{equation}
Therefore, we can get the belief $b(h_{a,n}^{(p)})$ of $h_{a,n}^{(p)}$ as
\begin{equation}
    \label{eq:Hapbelief}
    \begin{split}
        b(h_{a,n}^{(p)})&=m^{\mathrm{BP}}_{f_{B^{(p)}_{n}}\to h_{a,n}^{(p)}}\left(h_{a,n}^{(p)}\right)m_{f_{g^{(p)}_{n}}\to h_{a,n}^{(p)}}^{\mathrm{BP}}(h_{a,n}^{(p)})
        \\
        &=B_{h_n}^{(p)}\mathcal{CN}(h_{a,n}^{(p)},\hat{\mu}_{\mathrm{L},n}^{(p)},\hat{\varsigma}_{\mathrm{L},n}^{(p)})
        +(1-B_{h_n}^{(p)})\mathcal{CN}(h_{a,n}^{(p)},\hat{\mu}_{\mathrm{S},n}^{(p)},\hat{\varsigma}_{\mathrm{S}}^{(p)}),
    \end{split}
\end{equation}
where the parameters $B_{h_n}^{(p)},\{\hat{\varsigma}_{\mathrm{L},n}^{(p)},\hat{\mu}_{\mathrm{L},n}^{(p)}\}, \{\hat{\varsigma}_{\mathrm{S}}^{(p)},\hat{\mu}_{\mathrm{S},n}^{(p)}\}$ are calculated again through \eqref{eq:HSconstant},~\eqref{eq:Largemean} and~\eqref{eq:Smallmean}, respectively. However, here, the parameters $\hat{\epsilon}^{(p)}_n,\hat{\eta}^{(p)}_n,\hat{\alpha}^{(p)},\hat{\beta}^{(p)}$ used in \eqref{eq:HSconstant},~\eqref{eq:Largemean} and~\eqref{eq:Smallmean} are updated by~\eqref{eq:epcUpdate} and~\eqref{eq:alphaUpdate}. 

Finally, the output of module B is the mean and variance of the belief $b(h_{a,n}^{(p)})$ as
\begin{equation}
    \label{eq:HBpost}
    \begin{split}
        h_{\mathrm{B},n}^{\mathrm{post}(p)}=\mathbb{E}\left[h_{a,n}^{(p)}\right]=B_{h_n}^{(p)}\hat{\mu}_{\mathrm{L},n}^{(p)}+(1-B_{h_n}^{(p)})\hat{\mu}_{\mathrm{S},n}^{(p)},
    \end{split}
\end{equation}
\begin{equation}
    \label{eq:VBpost}
    \begin{split}
        v_{\mathrm{B}}^{\mathrm{post}(p)}=\frac{1}{N} \sum_{n=1}^{N} \mathbb{V} \mathrm{ar}\left(h_{a,n}^{(p)}\right)=\frac{1}{N} \sum_{n=1}^{N}\left\{   B_{h_n}^{(p)}(|\hat{\mu}_{\mathrm{L},n}^{(p)}|^2+\hat{\varsigma }_{\mathrm{L},n}^{(p)})+\right. 
        \\
        \left.(1-B_{h_n}^{(p)})(|\hat{\mu}_{\mathrm{S},n}^{(p)}|^2+\hat{\varsigma }_{\mathrm{S}}^{(p)})-|h_{\mathrm{B},n}^{\mathrm{post}(p)}|^2\right\}.
    \end{split}
\end{equation}
\subsection{Scheduling of the Messages}
The factors in Fig.~\ref{fig:systemfactor} are very densely connected and thus there are a multitude of different options for message scheduling. We summarize our schedule and the corresponding message computations in \textbf{Algorithm~\ref{alg:Hybrid}}.
As shown in \textbf{Algorithm~\ref{alg:Hybrid}}, some variables are initialized before the iterative process. The detailed initialization values of the parameters are given in the next Section. The \emph{Part~1} messages are first calculated in parallel for each subcarrier based on the extrinsic messages from module A. When the messages over all the subcarriers reach the Markov chain, the \emph{Part~2} and \emph{Part~3} messages are then updated in sequence, respectively. After that, we can update the hyperparameters $p_{10}$ and $p_{01}$ in the \emph{Part~4}. Finally, we can get more accurate \emph{Part~5} messages outgoing the Markov chain by updating the \emph{Part~2} and \emph{Part~3} messages again based on the updated hyperparameters $p_{10}$ and $p_{01}$ from the \emph{Part~4}. 

\begin{algorithm}[ht]
    \DontPrintSemicolon
    \SetNoFillComment
    \caption{HMP-TSGM-LVD algorithm}
    \label{alg:Hybrid}
    \footnotesize
    \KwInput{$\mathrm{Module~A~extrinsic~messages}~\boldsymbol{h}_{\mathrm{B}}^{\mathrm{pri}(p)},v_{\mathrm{B}}^{\mathrm{pri}(p)},~\forall p,\mathrm{~Maximum~iterations}~T.$}
    \KwOutput{$\mathrm{Module~B~channel~estimation~matrix~}\boldsymbol{h}_{\mathrm{B}}^{\mathrm{post}(p)},\forall p.$}
    \SetKwInput{KwInitialize}{Initialize}
    \KwInitialize{$\mathrm{Prior~distribution~parameters:~}\epsilon_n^{(p)},\eta_n^{(p)},\alpha^{(p)},\beta^{(p)},\forall n,\forall p;~e,f,c,d.$
    \\
    \hspace*{0.52in}  
    $\mathrm{Belief~parameters:~}\hat{\epsilon}_n^{(p)},\hat{\eta}_n^{(p)},\hat{\alpha}^{(p)},\hat{\beta}^{(p)},\forall n,\forall p;~\hat{e},\hat{f},\hat{c},\hat{d}$, and $\lambda^{\Uparrow }_N=1/2.$
    }
    
    \While{ convergence == FALSE or the iteration number less than T}{
    
    \tcp*[l]{Part 1 - Right part messages}
    $\forall p,\forall n$: update ${\rm{\mathord{\buildrel{\lower3pt\hbox{$\scriptscriptstyle\rightharpoonup$}} \over \pi}}}_n^{(p)}$ by \eqref{eq:phiright}\;
    
    \tcp*[l]{Part 2 - Downward messages}
   update $\lambda^{\downarrow }_1$ and $\lambda^{\Downarrow }_1$ by \eqref{eq:lamdadownMS1} and \eqref{eq:lamdadownNSn}\;
   
    $\forall n\in[2:N]$: update $\lambda^{\downarrow }_n$ and $\lambda^{\Downarrow}_n$ by \eqref{eq:lamdadownMSn} and \eqref{eq:lamdadownNSn}\;
    
    \tcp*[l]{Part 3 - Upward messages}
    $\forall n\in[N-1:1]$: update $\lambda^{\Uparrow }_n$ and $\lambda^{\uparrow }_n$ by \eqref{eq:lamdaupNSn} and \eqref{eq:lamdaupMSn}\;
    
    \tcp*[l]{Part 4 - Transition probability parameters update messages}
    update $B_{s_1}$ by \eqref{eq:Bs1}\;
    $\forall n\in[2:N]$: update $B_{n,n-1}^{00},B_{n,n-1}^{01},B_{n,n-1}^{10},B_{n,n-1}^{11},\rho_{n,n-1}$ by \eqref{eq:Bcomsn}\;
    update $\hat{e},\hat{f}~\mathrm{and}~\hat{c},\hat{d}$ by \eqref{eq:updateef} and \eqref{eq:updatecd}\;
    perform step 2-4 again\;
    
    \tcp*[l]{Part 5 - Left part messages}
    $\forall p,\forall n$: update ${\rm{\mathord{\buildrel{\lower3pt\hbox{$\scriptscriptstyle\leftharpoonup$}} \over \pi}}}_n^{(p)}$ by \eqref{eq:phileft}\;
    $\forall p,\forall n$: update $B_{h_n,s_n}^{(p)}$ by \eqref{eq:HSconstant}\;
    $\forall p,\forall n$: update $\hat{\varsigma}_{\mathrm{L},n}^{(p)},\hat{\mu}_{\mathrm{L},n}^{(p)}~\mathrm{and}~\hat{\varsigma}_{\mathrm{S}}^{(p)},\hat{\mu}_{\mathrm{S},n}^{(p)}$ by \eqref{eq:Largemean} and \eqref{eq:Smallmean}\;
    update $\hat{\epsilon}^{(p)}_{n},\hat{\eta}^{(p)}_{n}~\mathrm{and}~\hat{\alpha}^{(p)},\hat{\beta}^{(p)}$ by \eqref{eq:epcUpdate} and \eqref{eq:alphaUpdate}\;
    
    $\forall p,\forall n$: update $B_{h_n}^{(p)}$ by \eqref{eq:HSconstant}\;
    
    $\forall p,\forall n$: update $\hat{\varsigma}_{\mathrm{L},n}^{(p)},\hat{\mu}_{\mathrm{L},n}^{(p)}~\mathrm{and}~\hat{\varsigma}_{\mathrm{S}}^{(p)},\hat{\mu}_{\mathrm{S},n}^{(p)} $ by \eqref{eq:Largemean} and \eqref{eq:Smallmean} again\;
    
    $\forall p,\forall n$: update $h_{\mathrm{B},n}^{\mathrm{post}(p)},v_{\mathrm{B}}^{\mathrm{post}(p)}$ by \eqref{eq:HBpost} and \eqref{eq:VBpost}\;
    }  
\end{algorithm}

\section{Simulation Results}
\label{sec:Sim}
In this section, we first develop the SE to accurately predict the performance of the proposed HMP-TSGM-LVD algorithm. Then, we compare the performance of various algorithms using the channel generated in Section II-A. Previous works~\cite{STCSMP,STCSFS,ChenSTCS} have demonstrated that STCS-FS performs better than Turbo-CS~\cite{TurboDFT}, OMP~\cite{OMP}, DSAMP~ \cite{Gao2016FDD}, EM-BG-AMP~\cite{EMBGAMP}, AMP-NNSPL-FD~\cite{AMPNNSPLFD}. 
\revise{Hence, as a benchmark, we employ the STCS-FS algorithm~\cite{STCSMP,STCSFS} which applies the BP-EM to calculate the messages and update the parameters. In order to provide a thorough comparison, we combine the STCS-FS algorithm with BG and TSGM prior probability models.} 
The specific algorithms are labeled as follows:
``STCS-FS-BG" and ``STCS-FS-TSGM" stand for STCS-FS algorithm with BG~\cite{STCSMP,STCSFS} and TSGM prior models~\cite{BICMOFDM}, respectively; ``HMP-BG", ``HMP-TSGM" and ``HMP-TSGM-LVD" denote the HMP algorithm with BG, TSGM and TSGM-LVD prior models, respectively. It is worth noting that ``HMP-TSGM-LVD" denotes the proposed Algorithm~\ref{alg:Hybrid} \revise{when using the proposed probability model}, while the HMP-TSGM uses the same Algorithm~\ref{alg:Hybrid} but it differs in the channel prior model. In particular, TSGM assume that the variances of the non-zero elements are the same. For this scheme, \eqref{eq:epcUpdate} can be replaced by
\[
\hat{\epsilon}^{(p)}=\epsilon^{(p)}+\sum_{n=1}^{N}B_{h_n,s_n},~~~\hat{\eta}^{(p)}=\eta^{(p)}+\sum_{n=1}^{N}B_{h_n,s_n}(|\hat{\mu}_{\mathrm{L},n}^{(p)}|^2+\hat{\varsigma }_{\mathrm{L},n}^{(p)})
\]
and the rows 12-16 of Algorithm~\ref{alg:Hybrid} changes accordingly.

\subsection{Parameter Settings}
We consider the massive MIMO-OFDM system in which the BS is equipped with $N = 256$ antennas serving a single antenna user. The BS applies $P = 32$ pilot subcarriers to transmit $M=103\approx 0.4N$ training sequences continuously. We generate different pilot matrices $\boldsymbol{A}^{(p)}$ for different subcarriers by PDFT-RP~\cite{STCSMP}. The detailed SCM parameters listed in Table \ref{tab:Parameter} are used for all the simulations, if not specified.
 \begin{table}[h]
	\centering
	\caption{Parameter settings for the SCM.}
	\label{tab:Parameter}  
	\begin{tabular}{cccc} 
		\hline\hline\noalign{\smallskip}	
		Parameter name& Value& Parameter name& Value\\
		\noalign{\smallskip}\hline\noalign{\smallskip}
		NumBsElements $N$& 256& Subcarriers number $K$& 512
		\\
	    NumMsElements& 1& Pilot subcarriers $P$& 32
	    \\
        BandWidth& 15MHz& Subcarriers spacing& 15KHz
        \\
        CenterFrequency& 2GHz& NumPaths& 6
        \\
        Scenarios& \multicolumn{3}{c}{Urban macro, Suburban macro}
        \\
		\noalign{\smallskip}\hline
	\end{tabular}
\end{table}
\revise{The parameters $\{\epsilon ^{(p)}_n=1,\eta^{(p)}_n=1,\forall n,p\}$, $\{\alpha^{(p)}=1,\beta^{(p)}=0.01,\forall p\}$, $\{e = 1,f = 1\}$ and $\{c = 1,d = 1\}$\footnote{\revise{According to \cite[chapter 6.2]{Betasetting}, we set the parameters as $e=f=c=d=1$ and the hyperparameters $p_{10},p_{01}$ are uniform distributed since we do not know the prior information about the Markov chain transition probabilities.}} of the hyperparameters $v_{\mathrm{L},n}^{(p)}$, $v_{\mathrm{S}}^{(p)}$, $p_{10}$ and $p_{01}$ are empirically set, respectively.
For the parameters setting of $v_{\mathrm{L},n}^{(p)}$, $v_{\mathrm{S}}^{(p)}$, we can select a threshold value $\rho$ of the channel data modulus squared $|h_{a,n}^{(p)}|^2$ to intercept the non-zero ($|h_{a,n}^{(p)}|^2>\rho$) and near-zero ($|h_{a,n}^{(p)}|^2\le \rho$) elements positions. Then, we calculate the average power of the channel non-zero and near-zero elements separately as the prior characteristics of the channel probability model variance based on $S=1000$ channel realizations. Here, we have heuristically set $\rho = 0.05$ and obtained the variance of the large and small Gaussians as $v_{\mathrm{L},n}^{(p)} = 1$ and $v_{\mathrm{S}}^{(p)} = 0.01$. According to~\cite{GuoSBL}, the shape parameter of the Gamma distribution is usually set to $\epsilon_n^{(p)}=\alpha^{(p)} = 1$. Further, based on the expectation of the Gamma distribution $v_{\mathrm{L},n}^{(p)} = \eta_n^{(p)}/\epsilon_n^{(p)}$ and $v_{\mathrm{S}}^{(p)} = \beta^{(p)}/\alpha^{(p)}$, we set the scale parameters to $\eta_n^{(p)} = 1,\beta^{(p)} = 0.01$ respectively. 
At the first iteration of the algorithm, the parameters of the beliefs with $v_{\mathrm{L},n}^{(p)}$, $v_{\mathrm{S}}^{(p)}$, $p_{10}$ and $p_{01}$ are also empirically initialized as the value of the prior parameter values, i.e., $\{\hat{\epsilon}_n^{(p)}=\epsilon_n^{(p)},\hat{\eta}_n^{(p)}=\eta_n^{(p)},\forall n,\forall p;~\hat{\alpha}^{(p)}=\alpha^{(p)},\hat{\beta}^{(p)}=\beta^{(p)},\forall p;\hat{e}=e,\hat{f}=f,\hat{c}=c,\hat{d}=d\}$. The reason is that the algorithm starts with an initial value for the belief parameters, and then the algorithm will update the belief parameters during each iteration. In addition, it is a reasonable assumption to use the prior values for the initial setting when we do not know the belief parameters.}

All the curves are calculated by averaging 100 Monte Carlo simulations. Each simulation employs a new realization of the pilot matrix $\boldsymbol{A}^{(p)},\forall p$, the channel matrix $\boldsymbol{h}^{(p)}_a,\forall p$ and the AWGN matrix $\boldsymbol{\omega}^{(p)}_f,\forall p$. We use the NMSE $ =||\boldsymbol{\hat{H}}_a-\boldsymbol{H}_a||^2_2/ \left\|\boldsymbol{H}_a\right\|^2_2$ as a performance metric.

\subsection{State Evolution}
\begin{figure}
\begin{centering}
    \includegraphics[scale=0.65]{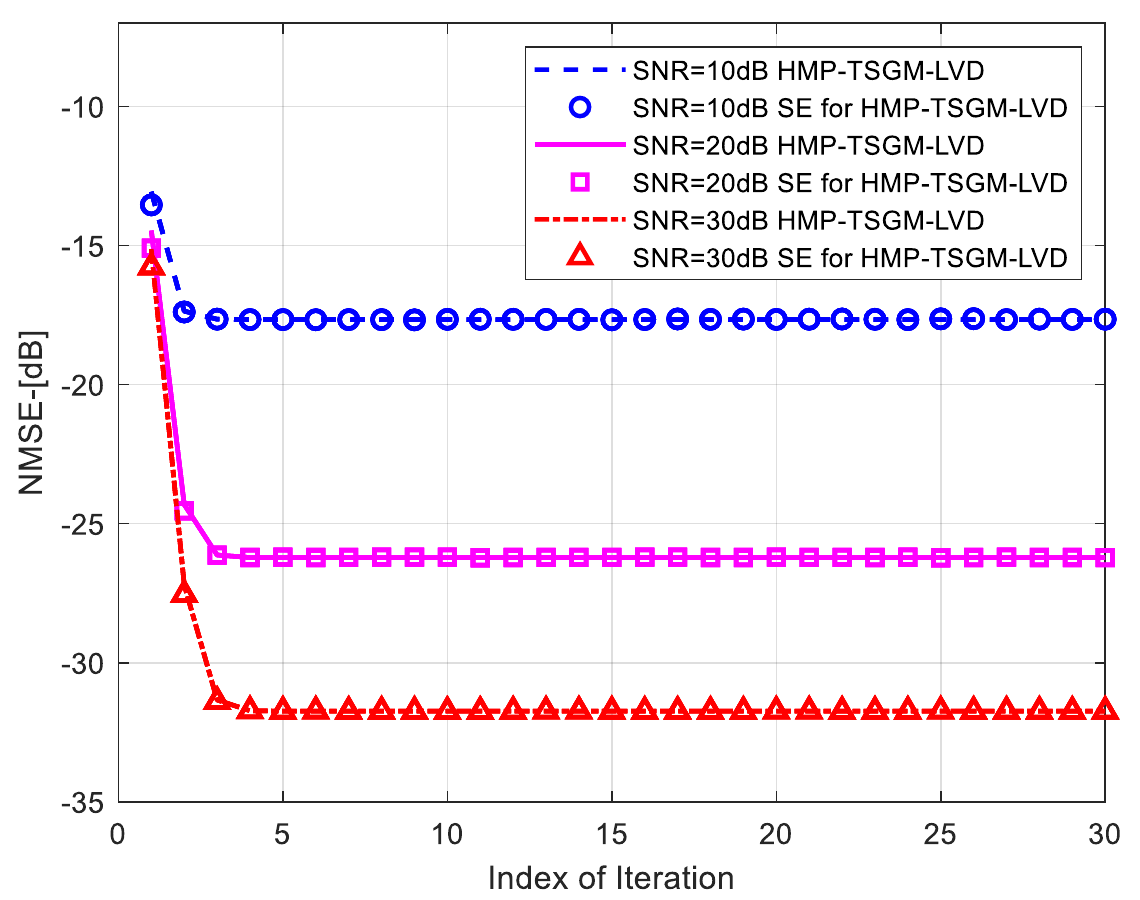}
    \caption{Comparison of SE and simulation results for HMP-TSGM-LVD under different SNR = 10dB, 20dB and 30dB. 
    Both experiments use parameters $N$ = 512, $M = 0.8N \approx  400$, and the results are averaged over 100 realizations.}
    \label{fig:SE}
\end{centering}
\end{figure}

The performance of HMP-TSGM-LVD can be characterized by simple scalar recursions called SE~\cite{Donoho,TurboDFT,MaTurbo,STCSFS}. We use the variance of the extrinsic messages $v_{\mathrm{A}}^{\mathrm{pri}(p)}$ and $v_{\mathrm{B}}^{\mathrm{pri}(p)}$ to measure the reliability of the channel estimator $\boldsymbol{h}_{\mathrm{A}}^{\mathrm{pri}(p)}$ and $\boldsymbol{h}_{\mathrm{B}}^{\mathrm{pri}(p)}$, respectively. We define
\begin{equation}
    \label{eq:SEvariable}
    v^{(p)}\triangleq  v_{\mathrm{A}}^{\mathrm{pri}(p)}~~\mathrm{and}~~
    \eta^{(p)} \triangleq \frac{1}{v_{\mathrm{B}}^{\mathrm{pri}(p)}}.
\end{equation}
Moreover, we define $mmse(\eta^{(p)})$ as the minimum mean squared error (MMSE) of the sparse signal estimation
given an AWGN observation, i.e.,
\begin{equation}
    \label{eq:mmse}
    mmse(\eta^{(p)}) =\mathbb{E}\left[\left|h^{(p)}_{a,n}-\mathbb{E}\left[h^{(p)}_{a,n}|h^{(p)}_{a,n}+\xi\right]\right|^2\right],
\end{equation}
where $h^{(p)}_{a,n}$ is a sparse signal modeled as \eqref{eq:assumeB} and $\xi\sim \mathcal{CN}(0,\eta ^{(p)-1})$. Compared with \eqref{eq:VBpost}, we have
\begin{equation}
    \label{eq:mmsecompare}
   v_{\mathrm{B}}^{\mathrm{post}(p)}=\frac{1}{N} \sum_{n=1}^{N} \mathbb{V} \mathrm{ar}\left(h_{a,n}^{(p)}|h_{\mathrm{B},n}^{\mathrm{pri} (p)}\right)\longrightarrow mmse(\eta^{(p)} ).
\end{equation}
Therefore, the SE of the proposed HMP-TSGM-LVD algorithm is characterized by
\begin{subequations}
\label{eq:SE}
\begin{align}
    \label{eq:SElikelihood}
    \eta^{(p)}_{t+1}=&~\varphi (v^{(p)}_t) =\frac{1}{\frac{N}{M}\cdot (v^{(p)}_t+\sigma ^2)-v^{(p)}_t},
    \\
     \label{eq:SEpost}
    \frac{1}{v^{(p)}_{t+1}}=&~\phi ( \eta^{(p)}_{t+1})=\frac{1}{mmse(\eta^{(p)}_{t+1})} - \eta^{(p)}_{t+1},
\end{align}
\end{subequations}
where the subscript $t$ and $t+1$ indicate the iteration indices, and the superscript $(p)$ indicates the pilot subcarrier index. For the detailed derivation of the state evolution in~\eqref{eq:SE} please refers to~\cite{TurboDFT}. 

Fig.~\ref{fig:SE} illustrates the NMSE performances of the HMP-TSGM-LVD algorithm proposed in this article, together with the predictions given by the SE~\eqref{eq:SE}. We find that HMP-TSGM-LVD convergence can be accurately predicted by the SE, which indicates the effectiveness of the proposed algorithm.

\subsection{Computational Complexity}
Our proposed HMP-TSGM-LVD algorithm has no matrix inverse operation and the complexity is mainly concentrated on matrix multiplication operations. The sensing matrix is chosen as a partial DFT matrix, which means the matrix multiplication can be substituted by FFT.
In this way, the complexity of initializing the sensing matrix $\boldsymbol{A}^{(p)},1\le p\le P$ is $\mathcal{O}(NP\mathrm{log}N)$. The complexity of each iteration attributes to the multiplications computing in line 2, 10-12, 14-16, of which the complexity is approximately $\mathcal{O}(NP)$. Therefore, the overall complexity is $\mathcal{O}(NP\mathrm{log}N+NP)$. This per-iteration complexity is the same as STCS-FS \cite{STCSMP,STCSFS}, but we will further show that HMP-TSGM-LVD algorithm can achieve better estimation performance than STCS-FS.

\subsection{Performance Comparisons}
\begin{figure}
\centering
    \begin{subfigure}[t]{.49\textwidth}
        \centering
        \includegraphics[scale=0.6]{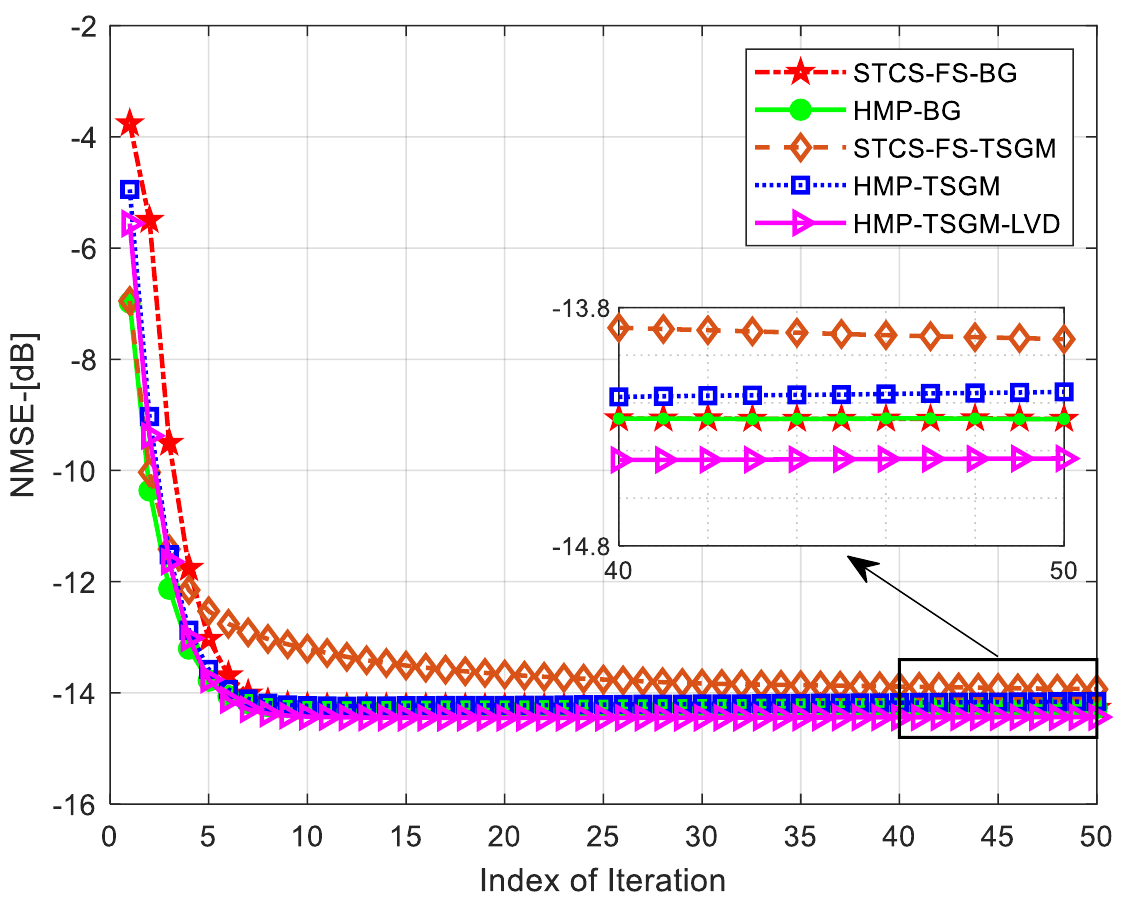}
        \caption{\small{Urban macro.}}\label{fig:Urbanmacro15}
    \end{subfigure}
    \hfill
    \begin{subfigure}[t]{.49\textwidth}
        \centering
        \includegraphics[scale=0.6]{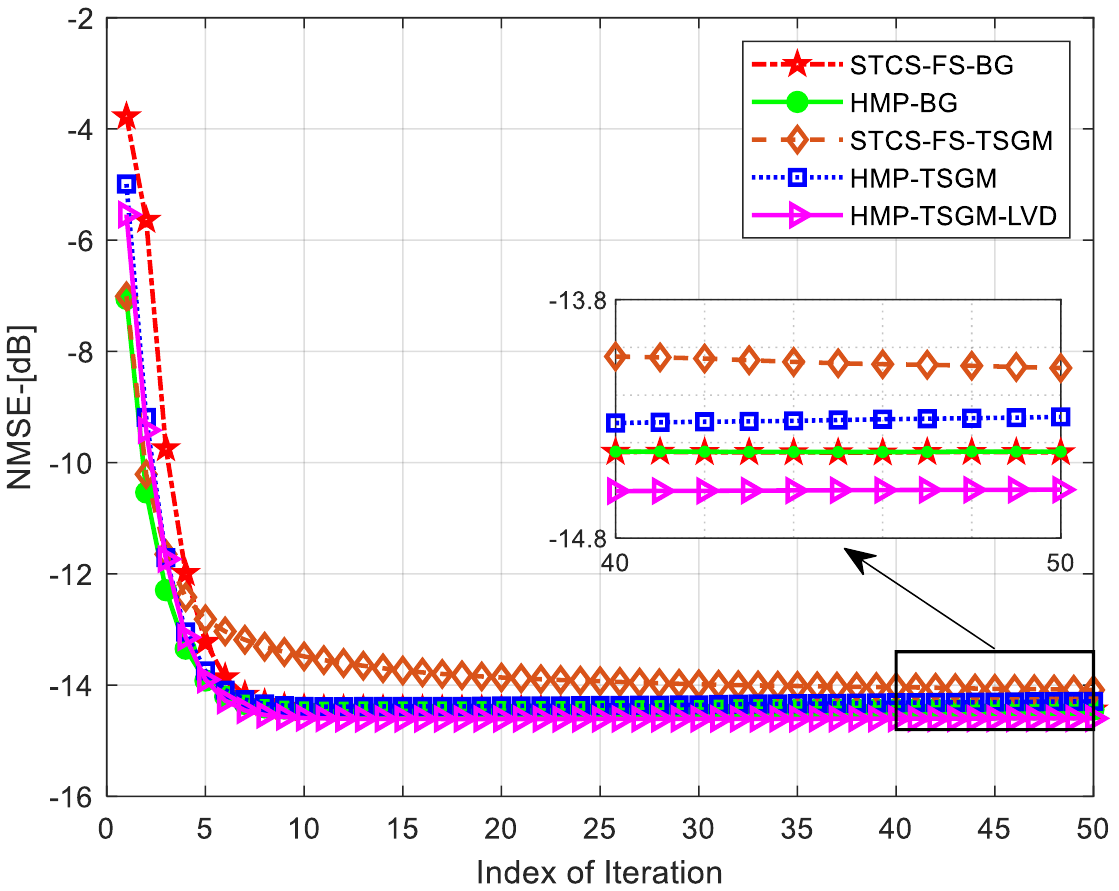}
        \caption{\small{Suburban macro.}}\label{fig:Suburbanmacro15}
    \end{subfigure}
    \\
    \begin{subfigure}[t]{.49\textwidth}
        \centering
        \includegraphics[scale=0.6]{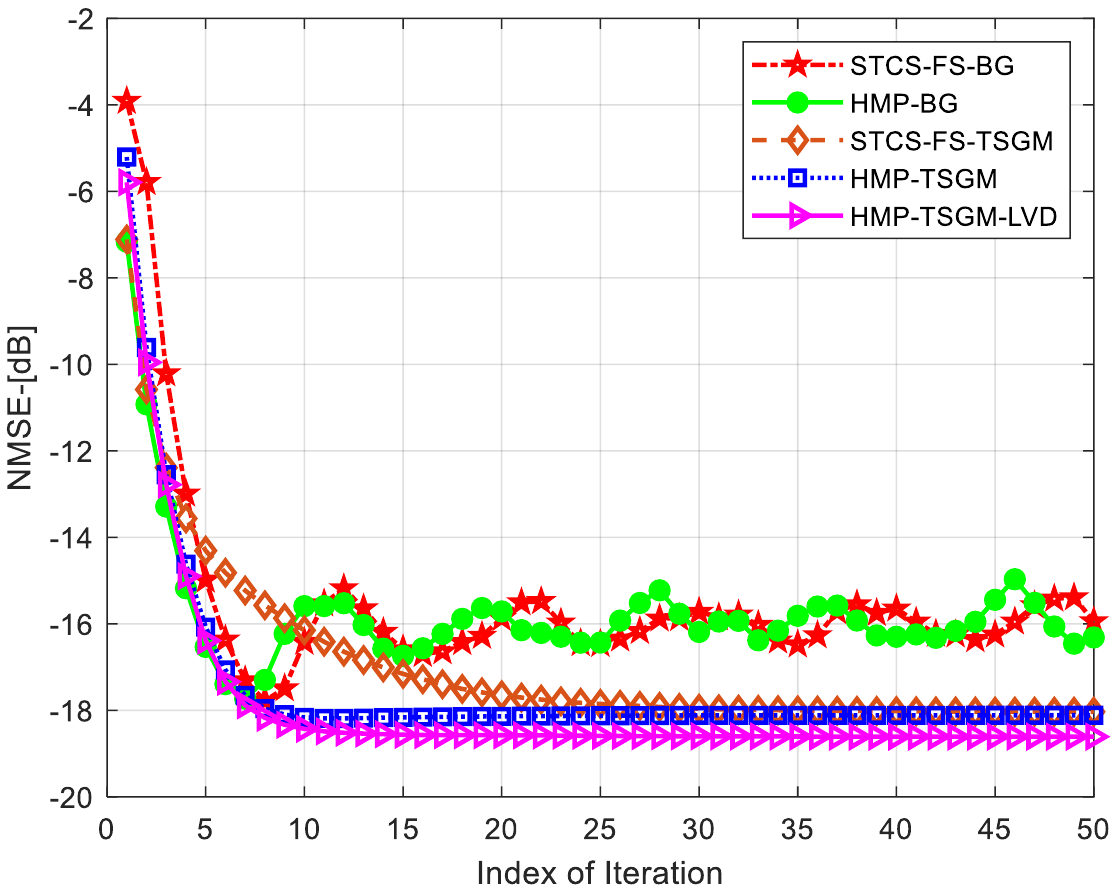}
        \caption{\small{Urban macro.}}\label{fig:Urbanmacro30}
    \end{subfigure}
    \hfill
    \begin{subfigure}[t]{.49\textwidth}
        \centering
        \includegraphics[scale=0.6]{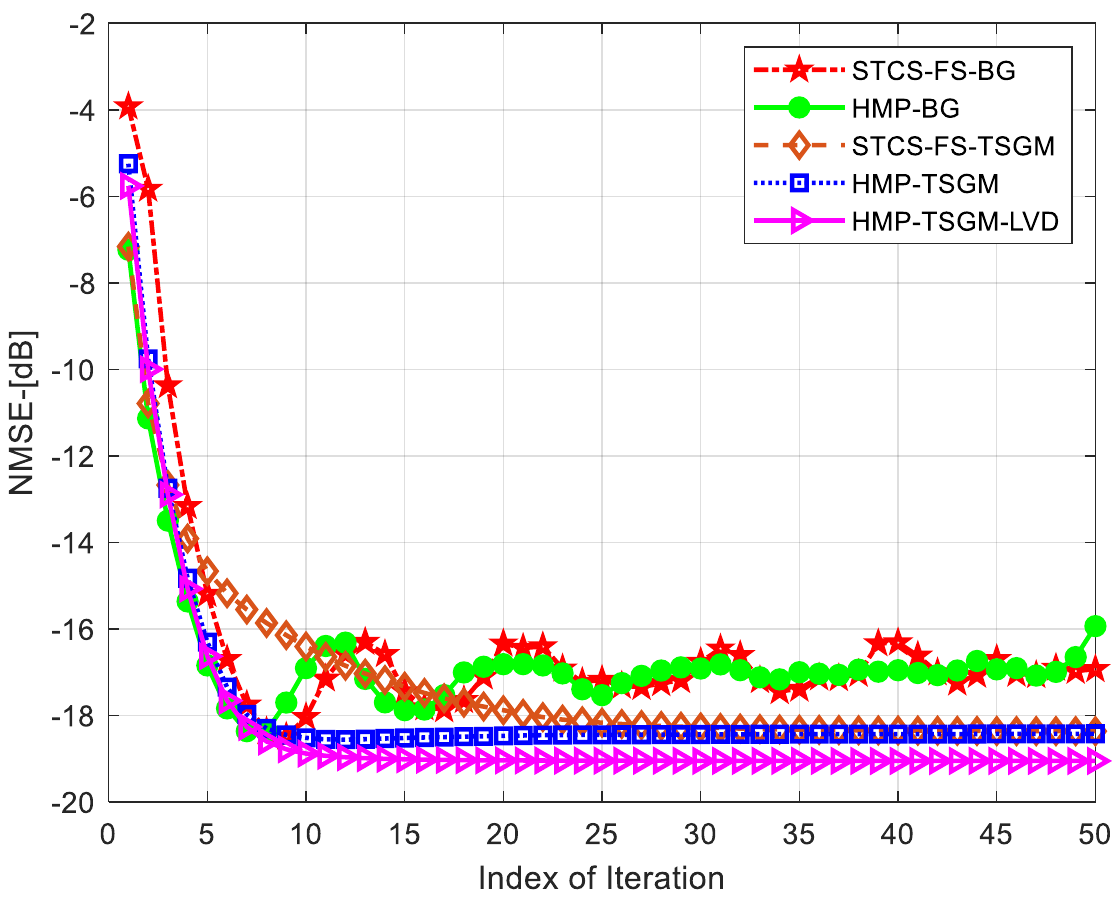}
        \caption{\small{Suburban macro.}}\label{fig:Suburbanmacro30}
    \end{subfigure}
\centering
\caption{NMSE performance versus iteration index, where the parameter $M$ = 103 is used for all subfigures. Subfigure (a), (b) expresses the experimental results for SNR = 15dB in two scenarios, while subfigure (c), (d) are under SNR = 30dB in two scenarios.}
\label{fig:Iter}
\end{figure}

We compare the convergence curves of the NMSE as a function of the number of iterations in Fig.~\ref{fig:Iter}. We perform two sets of experiments with signal-to-noise ratio (SNR) = 15dB and SNR = 30dB in the two SCM scenarios: urban macro, and suburban macro. These two scenarios are classified according to the location of the base station and the users~\cite{SCM}. As shown in Figs.~\ref{fig:Iter} (a) and (b), we observe that the performance of the ``HMP-TSGM'', ``HMP-TSGM-LVD'' and ``STCS-FS-BG'' is consistent when SNR = 15dB. However, Figs.~\ref{fig:Iter} (c) and (d) show that ``HMP-TSGM-LVD'' achieves better performance than other algorithms based on various prior models. This demonstrates that the TSGM-LVD model is able to capture the massive MIMO-OFDM channel characteristics more accurately than the BG or TSGM. In addition, with the same BG or TSGM prior models, HMP based algorithms converge faster than STCS-FS based algorithms while achieving the same performance.
%
It is worth noting that ``HMP-BG'' and ``STCS-FS-BG'' exhibit the ``oscillation" phenomenon as shown in Figs.~\ref{fig:Iter} (c), (d). The reason for this phenomenon is that, when the SNR is greater than 20 dB, the noise power is less than the power of the near-zero element of the channel. In this case, the channel near-zero elements can be regarded as the main influencing factor for CE, i.e., the BG prior model can not accurately characterize the near-zero elements of the channel.

\begin{figure}
\begin{centering}
    \begin{subfigure}[t]{.49\textwidth}
        \centering
        \includegraphics[scale=0.6]{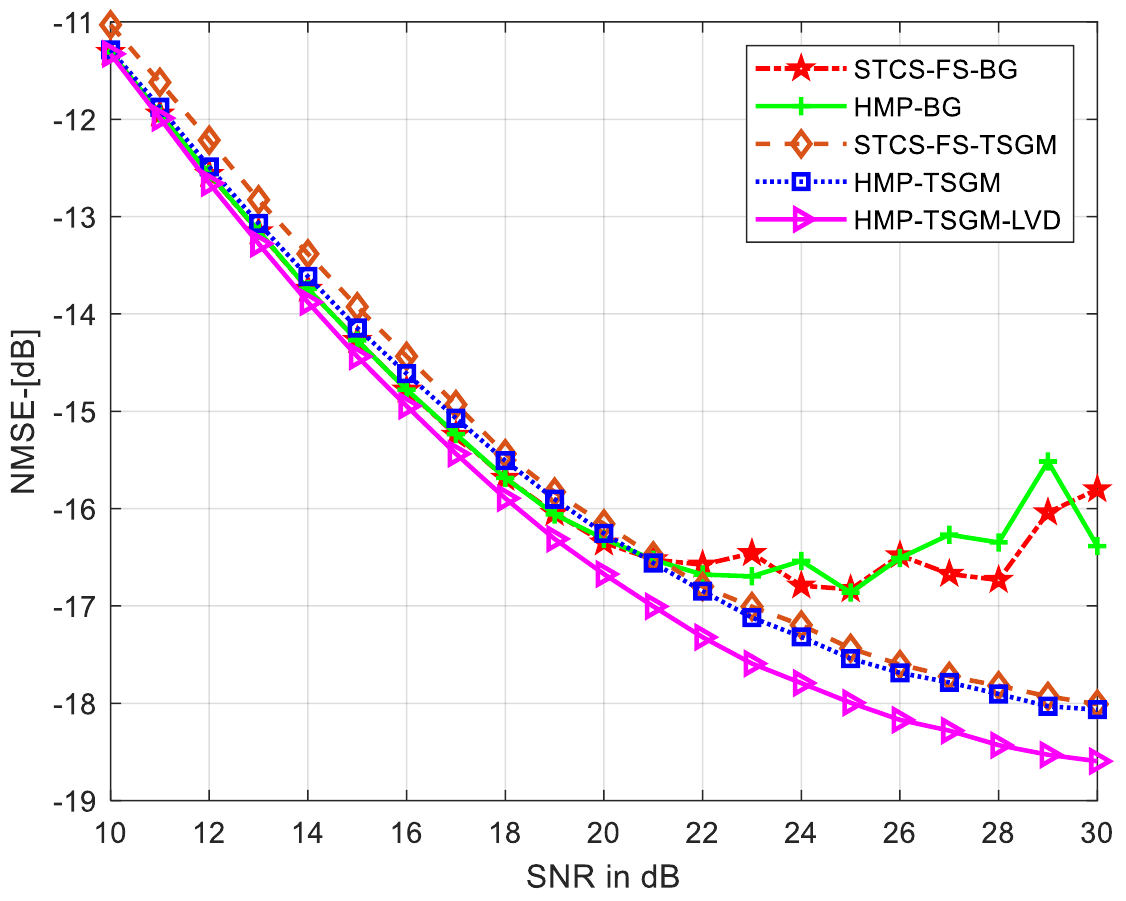}
        \caption{\small{Urban macro.}}\label{fig:SNRUrban}
    \end{subfigure}
    \hfill
    \begin{subfigure}[t]{.49\textwidth}
        \centering
        \includegraphics[scale=0.6]{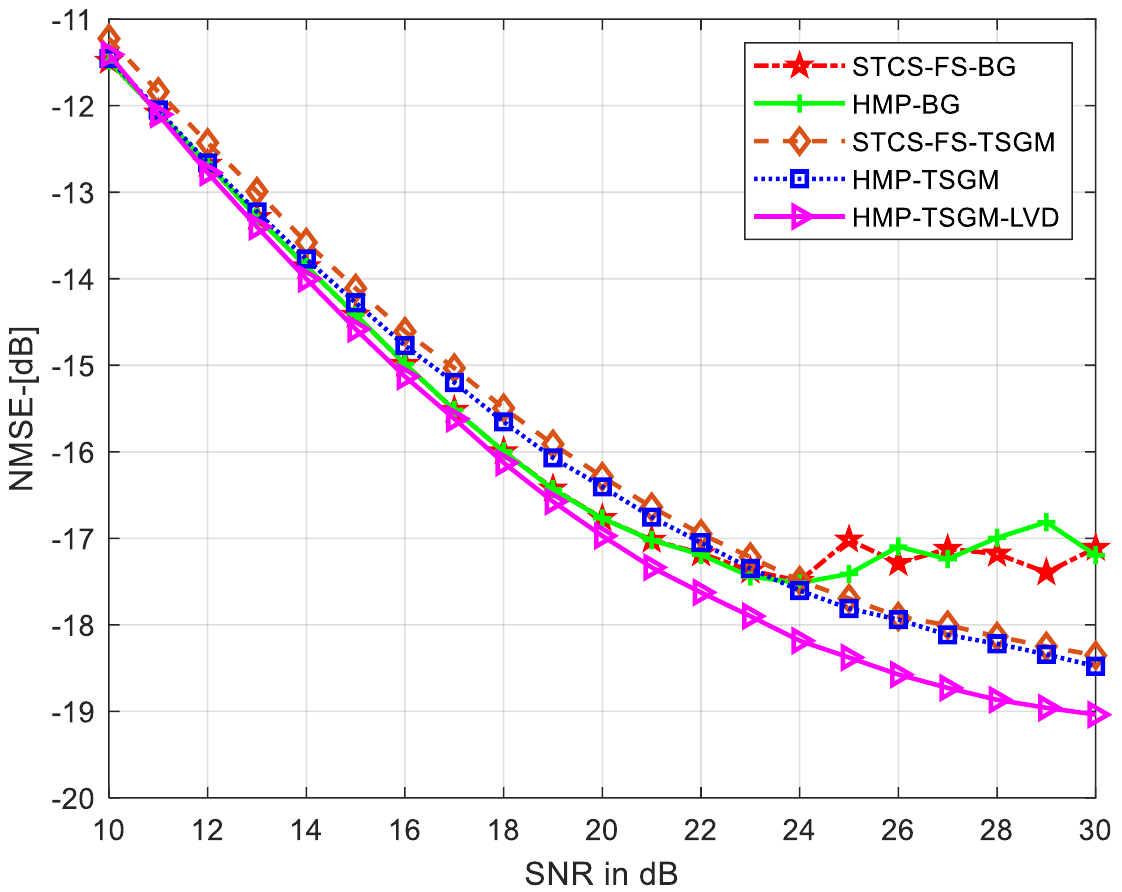}
        \caption{\small{Suburban macro.}}\label{fig:SNRSuburban}
    \end{subfigure}
\caption{NMSE performance with different SNRs. The system parameters are same in the subfigures with $N$ = 256, $M$ = 103, $P$ = 32.}
\label{fig:SNR}
\end{centering}
\vspace{-0.6cm}
\end{figure}

We compare the average NMSE performance as a function of the SNR in Fig.~\ref{fig:SNR}. We can observe that in Figs.~\ref{fig:SNR} (a) and (b), the proposed ``HMP-TSGM-LVD'' considerably outperforms the other algorithms within high SNR regime, i.e, SNR $\in [15:30]$ dB, while achieves the same performance for SNR $\in [5:15]$ dB. Therefore, the proposed TSGM-LVD prior model is proved to be more suitable to represent the massive MIMO-OFDM channel with different channel conditions.

\begin{figure}
\begin{centering}
        \begin{subfigure}[t]{.49\textwidth}
        \centering
        \includegraphics[scale=0.6]{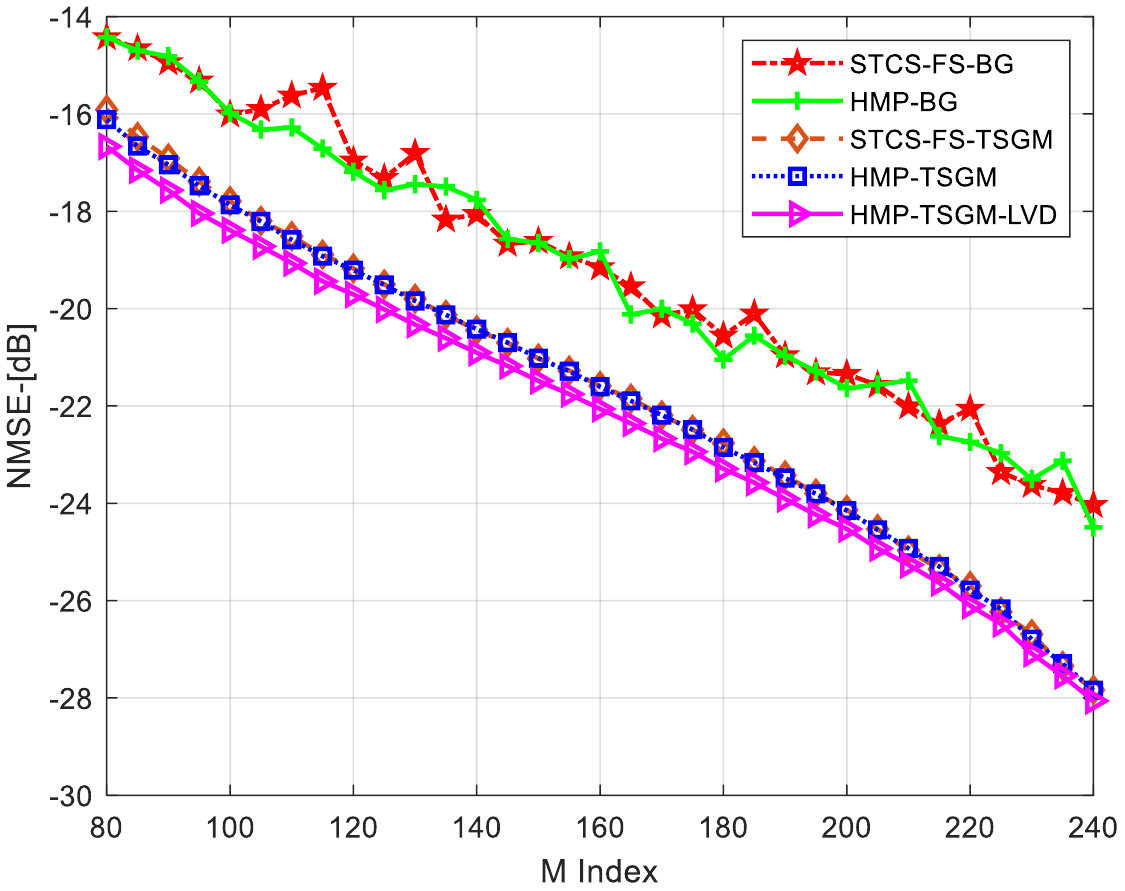}
        \caption{\small{ Urban macro.}}\label{fig:MUrbanmacro}
    \end{subfigure}
    \hfill
    \begin{subfigure}[t]{.49\textwidth}
        \centering
        \includegraphics[scale=0.6]{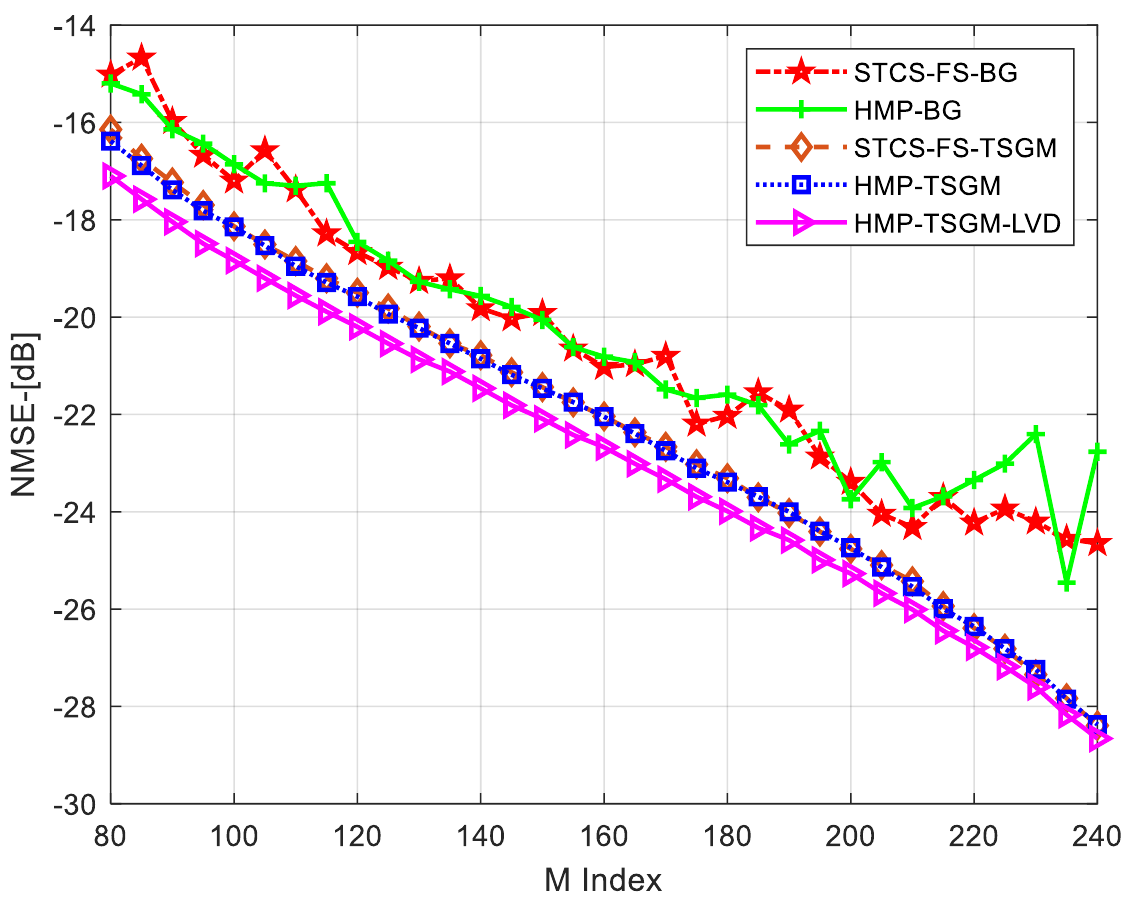}
        \caption{\small{Suburban macro.}}\label{fig:MSuburban}
    \end{subfigure}
\caption{NMSE performance versus the number of pilots $M \in [80:240]$. The system parameter is same in all the subfigures: SNR = 30 dB.}
\label{fig:M}
\end{centering}
\end{figure}

Finally, Fig.~\ref{fig:M} compares the average NMSE performance with a wide range of pilot numbers. Figs.~\ref{fig:M} (a) and (b) both demonstrate that ``HMP-TSGM-LVD'' performs better than any other algorithms regardless of the number of pilots employed. We remark that, under the same experimental conditions, TSGM-LVD model requires fewer pilots to achieve the same performance compared with the BG model. \revise{Nevertheless, the performance of HMP-TSGM is consistent with STCS-FS-TSGM, but the former converges faster than the latter proving the effectiveness of our approach.}

\section{Conclusion}
\label{sec:Conclusion}

In this article, we investigate the structured Turbo framework channel estimation problem in a massive MIMO-OFDM system. 
To exploit the sparsity structure of the channel in the AF domain, we proposed the TSGM-LVD probability model.
To solve the channel estimation problem in a practical communication system, we proposed a new method to derive the HMP rule, which is able to solve the joint product summation operations at the same factor node. 
By mixing the prior model and the HMP rule, we designed the HMP-TSGM-LVD channel estimation algorithm. In the simulations, we tested the approach in different SCM scenarios. It was demonstrated that the TSGM-LVD prior model fit the massive MIMO-OFDM channel characteristics better than other models in the literature. Moreover, we showed that the proposed algorithm converges faster and achieves better NMSE performance under a wide range of simulation settings, while having the same complexity as other state-of-the-art algorithms.

\begin{appendices}
\section{Derivation of Hybrid Message Passing Rule}
\label{sec:HMPproof}
\begin{figure}
\centering
    \includegraphics[scale=0.8]{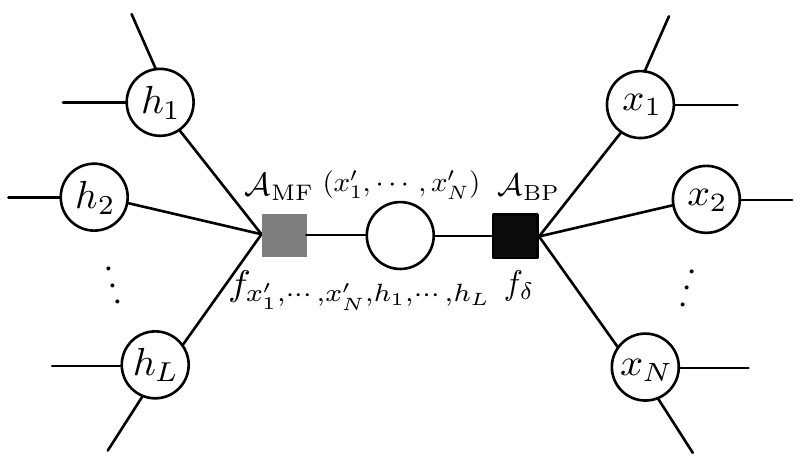}
    \caption{Stretched factor graph.}
    \label{fig:HMP_Proof}
\end{figure}

We consider the partial probability relationship as shown in Fig. \ref{fig:HMP}. Our target is to compute the messages $m_{f_a\to x_i}(x_i), i\in [1,N]$ and $m_{f_a\to h_l}(h_l), l\in [1,L]$. We can apply the combined BP-MF rule \cite{Merging} using the factor graph stretching approach \cite{BPMFstretched} to complete messages calculation. We modify the factor graph in Fig. \ref{fig:HMP} by adding hard constraint factors $f_{\delta}\triangleq \delta({x}'_1-x_1)\cdots\delta({x}'_N-x_N)$ with a new combined variable $({x}'_1,\cdots,{x}'_N)$. The new factor graph, shown in Fig. \ref{fig:HMP_Proof}, looks like a stretched version
of the graph in Fig. \ref{fig:HMP}. In the new graph, we first group the factor nodes into two sets: $\mathcal{A}_{\mathrm{BP}}=\left\{f_{\delta}\right\}$ and $\mathcal{A}_{\mathrm{MF}}=\left\{f_{new}\right\}$\footnote{For simplicity of writing, we define $f_{new} \triangleq  f_{{x}'_1,\cdots,{x}'_N,h_1,\cdots,h_L}$.}. For factor nodes in the BP part, we calculate the messages to neighboring variable nodes using \eqref{eq:comBP}, and send extrinsic messages. For factor nodes in the MF part, messages to neighboring variable nodes are computed by \eqref{eq:comMF}, and beliefs are passed. The messages calculation procedure is presented in the following.
%
%
The message from factor node $f_{\delta}$ to variable node $({x}'_1,\cdots,{x}'_N)$ use \eqref{eq:comBP} as
\begin{equation}
    \label{eq:deltatoxs}
    \begin{split}
        m_{f_{\delta }\to {x}'_1,\cdots,{x}'_N}({x}'_1,\cdots,{x}'_N)&=\int \delta({x}'_1-x_1)\cdots\delta({x}'_N-x_N)n_{x_1\to f_{\delta }}(x_1)\cdots n_{x_N\to f_{\delta }}(x_N)\mathrm{d}x_1\cdots \mathrm{d}x_N
        \\
        &=n_{x_1\to f_{\delta }}({x}'_1)\cdots n_{x_N\to f_{\delta }}({x}'_N).
    \end{split}
\end{equation}
Then, message $m_{f_{new}\to h_l}(h_l)$ from $f_{new}$ to $h_l$ is obtained by \eqref{eq:comMF},
\begin{equation}
    \label{eq:ftolamda}
    \begin{split}
        m_{f_{new}\to {h}_l}({h}_l)=\mathrm{exp}\left\{\left\langle \mathrm{ln}f_{new}\right \rangle _{b({x}'_1,\cdots,{x}'_N)\prod_{j\in \mathcal{I}_{\mathrm{MF}}\setminus l}b(h_j)}\right \}, \forall l\in \mathcal{I}_{\mathrm{MF}},
    \end{split}
\end{equation}
where the combined belief of $({x}'_1,\cdots,{x}'_N)$ reads
\begin{equation}
    \label{eq:beliefxsHMP}
    \begin{split}
        b({x}'_1,\cdots,{x}'_N)=m_{f_{new}\to {x}'_1,\cdots,{x}'_N}({x}'_1,\cdots,{x}'_N)n_{x_1\to f_{\delta }}({x}'_1)\cdots n_{x_N\to f_{\delta }}({x}'_N).
    \end{split}
\end{equation}
The message $m_{f_{new}\to {x}'_1,\cdots,{x}'_N}({x}'_1,\cdots,{x}'_N)$ in \eqref{eq:beliefxsHMP} is computed with \eqref{eq:comMF}, obtaining
\begin{equation}
    \label{eq:ftoxs}
    \begin{split}
        m_{f_{new}\to {x}'_1,\cdots,{x}'_N}({x}'_1,\cdots,{x}'_N)=\mathrm{exp}\left\{\left\langle \mathrm{ln}f_{new}\right \rangle _{\prod_{l\in \mathcal{I}_{\mathrm{MF}}}b(h_l)}\right \},
    \end{split}
\end{equation}
where the beliefs $b(h_l),l\in [1,L]$ are updated by MF rule. Meanwhile, $n_{{x}'_1,\cdots,{x}'_N\to f_{\delta }}({x}'_1,\cdots,{x}'_N)$ is same as \eqref{eq:ftoxs}, i.e.,
\begin{equation}
    \label{eq:nxstodelta}
        n_{{x}'_1,\cdots,{x}'_N\to f_{\delta }}({x}'_1,\cdots,{x}'_N)=m_{f_{new}\to {x}'_1,\cdots,{x}'_N}({x}'_1,\cdots,{x}'_N).
\end{equation}
Finally, with the message \eqref{eq:nxstodelta}, we can calculate the message $m_{f_{\delta }\to x_i}(x_i)$ by BP rule as
\begin{equation}
    \label{eq:deltatox}
        m_{f_{\delta }\to x_i}(x_i)
        =\int \mathrm{exp}\left\{\left\langle \mathrm{ln}f_{new}\right \rangle _{\prod_{l\in \mathcal{I}_{\mathrm{MF}}}b(h_l)}\right \}\prod_{j\in \mathcal{I}_{\mathrm{BP}}\setminus i}n_{x_j\to f_{\delta }}(x_j)\mathrm{d}x_j.
\end{equation}
We refer to \eqref{eq:ftolamda}, \eqref{eq:beliefxsHMP} and \eqref{eq:deltatox} are HMP rule, which is well-suited to mixed linear and non-linear scenarios, e.g., the channel probability model shown in \eqref{eq:Priormodel} in Section \ref{sec:clusteredsparse}.


\end{appendices}

\section*{Acknowledgement}
The authors thank Prof. Petar Popovski from Aalborg University, Denmark, for the insightful discussions and suggestions that greatly assisted this work. 

\bibliographystyle{IEEEtran}
\bibliography{IEEEabrv,reference}
\end{document}